# Spin-Polarized Antiferromagnetic Spintronics

*Zhenzhou Guo, Xiaotian Wang\*, Wenhong Wang, Gang Zhang, Xiaodong Zhou\*, Zhenxiang Cheng\**

Z. Guo, X. Wang, Z. Cheng

Institute for Superconducting and Electronic Materials, Faculty of Engineering and Information Sciences, University of Wollongong, Wollongong, New South Wales 2500, Australia.
\*E-mail: xiaotianw@uow.edu.au; cheng@uow.edu.au

W. Wang,

School of Electronic and Information Engineering, Tiangong University, Tianjin 300387, China

G. Zhang

Yangtze Delta Region Academy of Beijing Institute of Technology, Jiaxing 314000, China

X. Zhou

School of Physical Science and Technology, Tiangong University, Tianjin 300387, China
\*E-mail: xdzhou323@gmail.com

Keywords: spin polarization, altermagnets, noncollinear antiferromagnets, layered antiferromagnets, magnetic transport properties

Spin-polarized antiferromagnets (AFMs), including altermagnets, noncollinear AFMs, and two-dimensional layer-polarized AFMs, have emerged as transformative materials for next-generation spintronic and optoelectronic technologies. These systems uniquely combine spin-polarized electronic states with vanishing net magnetization, enabling ultrafast spin dynamics, high-density integration, and robustness against stray magnetic fields. Their unconventional symmetry-breaking mechanisms—governed by crystal symmetry, chiral spin textures, or interlayer potential control—give rise to emergent phenomena previously exclusive to ferromagnets: nonrelativistic spin-momentum locking, spontaneous anomalous transport phenomena, gate-tunable magneto-optical responses, and nonrelativistic spin-polarized current. This review systematically examines the fundamental principles linking symmetry, band topology, and transport properties across these material classes, synthesizing recent breakthroughs in both theory and experiment. We further identify critical challenges in achieving room-temperature functionality, scalable Néel vector control, and coherent spin-current manipulation, while outlining pathways to harness these materials for ultra-low-power memory, spin-logic architectures, and quantum information technologies.

## 1. Introduction

The discovery of electron spin in 1925 ignited a century-long quest to harness spin degrees of freedom, ultimately giving rise to the field of spintronics.[1–7] By exploiting the coupling between spin and charge/orbital dynamics, spintronics promises to overcome the fundamental limitations of charge-based electronics. Early breakthroughs—such as spin-polarized tunneling in ferromagnetic junctions and half-metallic ferromagnets (FMs)—established FMs as the cornerstone of spintronic devices.[8,9] FMs exhibit spin-splitting band structure due to broken time-reversal symmetry ($\mathcal{T}$), enabling many novel transport phenomena.[2–5] However, their finite net magnetization imposes intrinsic constraints: limited integration density due to stray fields, susceptibility to external magnetic perturbations, and speed limits imposed by GHz-range magnetization dynamics. Conventional antiferromagnets (AFMs) with their fully compensated magnetic sublattices, inherently avoid these issues. Their vanishing net magnetization not only renders AFMs insensitive to external magnetic fields but also suppresses stray magnetic fields, ensuring that AFMs avoid magnetic interference in densely integrated spintronic architectures. Additionally, their inherently strong exchange interactions support THz-range precession frequencies, outperforming FMs by orders of magnitude.[6,10–14] Despite these advantages, traditional AFMs, such as CuMnAs, $Mn_2Au$, FeRh, and $MnBi_2Te_4$, are often constrained by combined $\mathcal{TS}$ symmetry ($\mathcal{S}$ being space inversion $\mathcal{P}$ or translation $t$), which enforces spin degeneracy across all wave vectors.[10,15–18] This symmetry restriction suppresses intrinsic spin polarization and anomalous transport properties, relegating AFMs to passive roles in early spintronic architectures.

To address this limitation, recent advances have shattered this paradigm through the discovery of spin-polarized AFMs—categorized into three groups based on their distinct strategies to break $\mathcal{TS}$ symmetry: altermagnets (AMs), noncollinear AFMs (ncl-AFMs), and two-dimensional (2D) layer-polarized AFMs (LP-AFMs) (**Figure 1**). AMs interconnect antiparallel spin sublattices through crystal-rotation symmetries rather than lattice translation or inversion as in conventional AFMs, generating alternating spin-split bands with *d*-, *g*-, or *i*- wave momentum dependence and spin-degenerate nodal topology without relativistic spin-orbit coupling.[19–21] Ncl-AFMs typically arrange spins in triangular or kagome lattices, creating chiral spin textures that break $\mathcal{TS}$ symmetry and induce Berry curvature-driven anomalous Hall effects (AHE).[22–24] LP-AFMs leverage interlayer potential gradients in 2D layered materials or heterostructures to break $\mathcal{TS}$ symmetry, enabling layer-selective spin polarization that can be tuned by multiple degrees of freedom, such as an applied electric field, stacking order or sliding ferroelectricity.[25–28] These materials uniquely amalgamate the virtues of FMs (spin-polarized currents) and AFMs (zero net moments, THz dynamics). Their symmetry-engineered band structures unlock phenomena once deemed exclusive to FMs: non-relativistic spin-momentum locking, $\mathcal{T}$-odd spin current, spontaneous AHE, anomalous Nernst effect (ANE), anomalous thermal Hall effect (ATHE), and magneto-optical effects (MOEs) (**Figure 1**).

Experimental realizations—from noncollinear bulk crystals to altermagnetic thin films and bilayer LP material—have validated their potential for high-density memory, ultrafast logic, and quantum coherent devices.

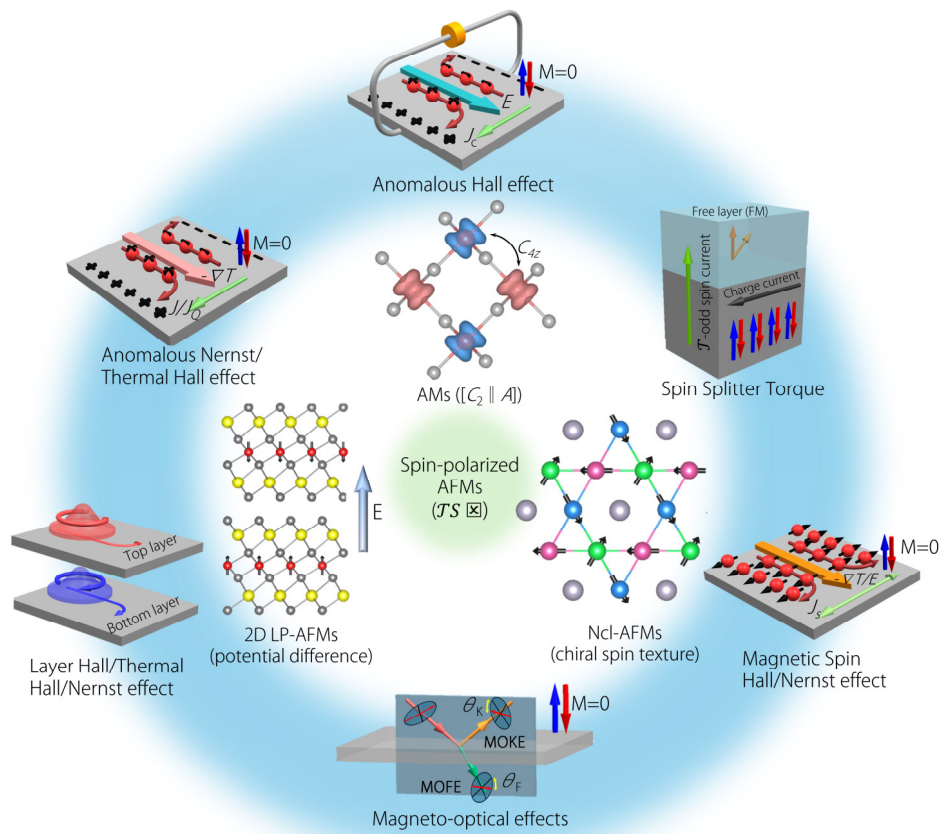

**Figure 1.** Schematics of three types of spin-polarized AFMs—AMs, ncl-AFMs, and LP-AFMs—highlighting their unique symmetry-engineered spin-polarized properties and the associated anomalous and spin transport effects.

In recent years, significant progress has been achieved in this area, driven by breakthroughs in fundamental theories and advancements in experimental techniques. Given that several review articles have systematically summarized various aspects of AFMs,[6,7,11,29] including the spin-polarized mechanism,[14,23,24,30–33] epitaxial growth and chemical synthesis methodologies,[34] electronic band structure and transport properties,[35–40] and the design of spintronic devices,[13,41–46] as well as other related topics,[47–49] we focus here on the latest research advances, providing a comprehensive overview from both theoretical and experimental perspectives, covering AMs, ncl-AFMs and LP-AFMs. We divide the three categories of spin-polarized AFMs into separate chapters, each dedicating a detailed discussion of their spin-polarization mechanisms, reviewing the theoretical predictions and experimental observations of their transport properties, and emphasizing the crucial role of antiferromagnetic crystal symmetry in shaping their transport characteristics. Furthermore, we highlight the challenges in material discovery, external field manipulation, and device integration, offering some thoughts on potential future research directions in this rapidly evolving field.

WILEY-VCH## 2. Altermagnets

### 2.1. Spin-momentum locking effect

Momentum-dependent spin splitting is a crucial phenomenon in modern condensed matter physics, as it bridges the gap between spin and momentum in electronic systems and could provide new avenues for efficient spin control and information transmission. In nonmagnetic systems, the most iconic examples are the Rashba and Dresselhaus effects, where spin-orbit coupling (SOC) interacts with broken inversion symmetry to generate momentum-dependent spin textures. However, such systems face inherent limitations: their splitting strength is constrained by SOC intensity and often requires external fields or engineered interfaces to break symmetry, restricting their practical utility. Intriguingly, researchers have realized that antiferromagnetic materials, long overlooked due to their zero net magnetization, can also exhibit pronounced momentum-dependent spin splitting even in the absence of SOC, opening up new possibilities for spintronic applications. [50–53]

In 2022, Šmejkal *et al.* proposed altermagnetism as a third fundamental magnetic order alongside ferromagnetism and conventional antiferromagnetism, completing the classification of spin collinear magnetic crystals via non-relativistic spin space group theory.[19–21] In FMs (including ferrimagnets), spin sublattices comprise either one spin sublattice or two opposite spin sublattices that cannot be connected by any symmetry operation. The inherent nonzero magnetization breaks $\mathcal{T}$ symmetry, enabling spin-splitting at arbitrary **k**-point in the Brillouin zone. As for conventional AFMs, they exhibit broken $\mathcal{T}$ symmetry but preserve Kramers spin degeneracy across all **k**-points due to two spin sublattices interconnected by $\mathcal{TS}$ symmetry. However, AMs feature spin sublattices linked by crystal-rotation symmetries (proper/improper, symmorphic/nonsymmorphic) rather than translation or inversion (**Figures 2A-B)**. This distinct configuration yields many extraordinary characteristics, including broken $\mathcal{T}$ symmetry, nonrelativistic alternating spin-splitting with anisotropic *d*-, *g*- or *i*-wave Fermi surfaces, and spin-degenerate nodal topology. Interestingly, the coupling between alternating magnetic order and ferroelectric or antiferroelectric properties can give rise to intriguing magnetoelectric effects, where the reversal of ferroelectric polarization can induce the switching of altermagnetic spin splitting.[54–58] The altermagnetic spin-splitting has been experimentally confirmed in several room-temperature altermagnetic materials like tetragonal $RuO_2$,[59] $KV_2Se_2O$,[60] and $Rb_{1-\delta}V_2Te_2O$,[61] as well as hexagonal $MnTe$[62–65] and $CrSb$.[66–69]

The key electronic properties in AMs can be derived by considering nonrelativistic spin group symmetry operations $[R_i\|R_j]$ on energy band $E(\mathbf{k},s)$, where the transformation on the left of the double vertical bar ($\|$) operates solely in spin space, and the transformation on the right operates in real space. For tetragonal $RuO_2$, $KV_2Se_2O$, and $Rb_{1-\delta}\ V_2Te_2O$, the $[C_2\|C_{4z}]$ operation transforms energy bands as $[C_2\|C_{4z}]E(k_x,k_y,k_z,s) = E(-k_y,k_x,k_z,-s)$, producing an unconventional *d*-wave magnetic phase (**Figures 2C-D**). Similarly, the $[C_2\|M_{x/y}]$ $[C_2\|M_{xy/\bar{x}y}]$) operations generate spin degenerate nodal surface on the $M_{x/y}$ ($M_{xy/\bar{x}y}$) invariant planes for RuO$_2$ (KV$_2$Se$_2$O and Rb$_{1-\delta}$V$_2$Te$_2$O).[21,70,71] In hexagonal MnTe and CrSb, $[C_2\|M_z]$ and $[C_2\|C_{6z}]$ operations yield spin-degenerate nodal planes along all high-symmetry lines, with $g$-wave spin-split bands emerging in non-symmetric regions of the Brillouin zone (**Figures 2E-H**).[62,63,67,72] Experimental studies corroborate these features: RuO$_2$ thin films exhibit strong $\mathcal{T}$ symmetry breaking via magnetic circular dichroism (**Figure 3A**),[59] while MnTe shows lifted Kramers spin degeneracy outside the nodal planes (**Figure 3B**).[62–65] CrSb similarly displays split bands near the Fermi level via spin-integrated soft X-ray angular-resolved photoemission spectroscopy (**Figure 3C**).[66–69] Notably, altermagnetic spin-splitting energy scales rival those of FMs (~eV scale), underscoring their significant potential for spintronic applications.[20]

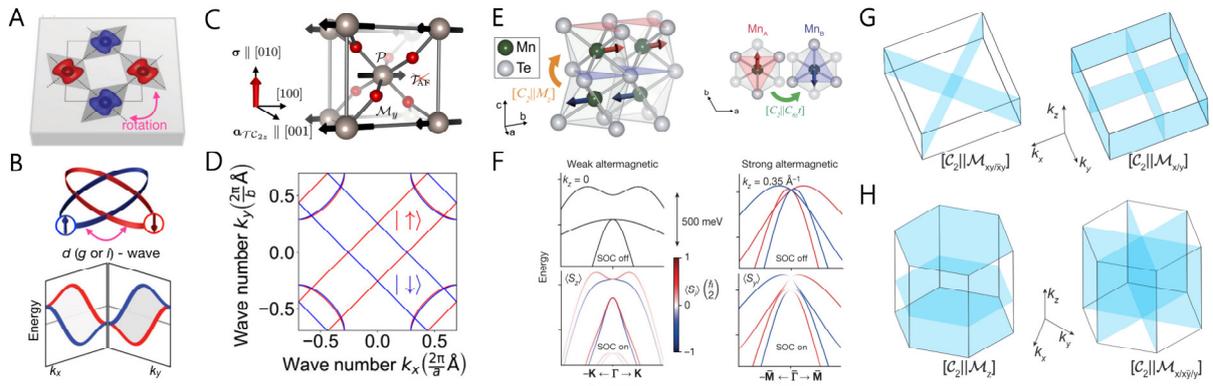

**Figure 2.** A) Altermagnetic model with opposite-spin sublattices linked by rotation. B) Corresponding anisotropic $d$ ($g$ or $i$)-wave spin-split Fermi surfaces and nonrelativistic alternating spin-splitting. Reproduced under the terms of the CC-BY 4.0 license.[21] Copyright 2022, The Authors, Published by American Physical Society. C) Unit cell of antiferromagnetic RuO$_2$ with the Néel vector aligned along the [100] axis. D) Fermi surface slice of RuO$_2$ at $k_z$=0 without SOC. Reproduced with permission.[70] Copyright 2020, American Association for the Advancement of Science. E) MnTe crystal structure with two opposite-spin sublattices (Mn$_A$ and Mn$_B$) connected by real-space mirror or sixfold rotation. Reproduced under the terms of the CC-BY 4.0 license.[62] Copyright 2024, The Authors, Published by American Physical Society. F) Electronic band structure of MnTe at $k_z$=0 along the Γ–K path and Γ–M path, respectively, with relativistic SOC turned off and on. Reproduced with permission.[63] Copyright 2024, Springer Nature Limited. G) Spin-degenerate nodal surfaces in tetragonal KV$_2$Se$_2$O and Rb$_{(1-\delta)}$V$_2$Te$_2$O (left panel) as well as in RuO$_2$ and MnF$_2$ (right panel). H) Similar to G) but in hexagonal MnTe and CrSb. Reproduced under the terms of the CC-BY 4.0 license.[72] Copyright 2025, The Authors, Published by arXiv.

To date, numerous materials have been theoretically predicted as altermagnets. High-throughput computational screening has identified a large number of candidate materials with spin-momentum locking band structures,[73,74] and a comprehensive list has been summarized in a recent review.[14] However, reduced symmetry in 2D systems often diminishes altermagnetic properties compared to three-dimensional (3D) bulk analogs,[75] leaving fewer confirmed 2D AMs, such as $V_2X_2O$ ($X$ = Se, Te)[76] and $Ca(CoN)_2$.[77] To expand the family of 2D AMs and facilitate their experimental exploration and spintronic applications, researchers have developed a series of methods that enable the preservation of non-relativistic spin group symmetries while breaking the $\mathcal{TS}$ symmetric operations of materials, thereby transforming conventional collinear AFMs into AMs. These strategies include applying external electric fields, chemical substitution of elements (e.g., "Janusization"), substrate effects, and supercell AMs.[78,79] Additionally, introducing altermagnetism in twisted bilayer structures of arbitrary 2D van der Waals materials through specific two-fold rotational symmetry operations has been proposed as a promising approach.[80,81] Furthermore, Pan et al. developed a general stacking theory for constructing AMs in bilayer systems through tailored stacking operations, enabling the design of diverse anisotropic Fermi surfaces.[82] These findings not only deepen our understanding of 2D altermagnetic phenomena but also provide a solid foundation for exploring their applications in next-generation spintronic devices.

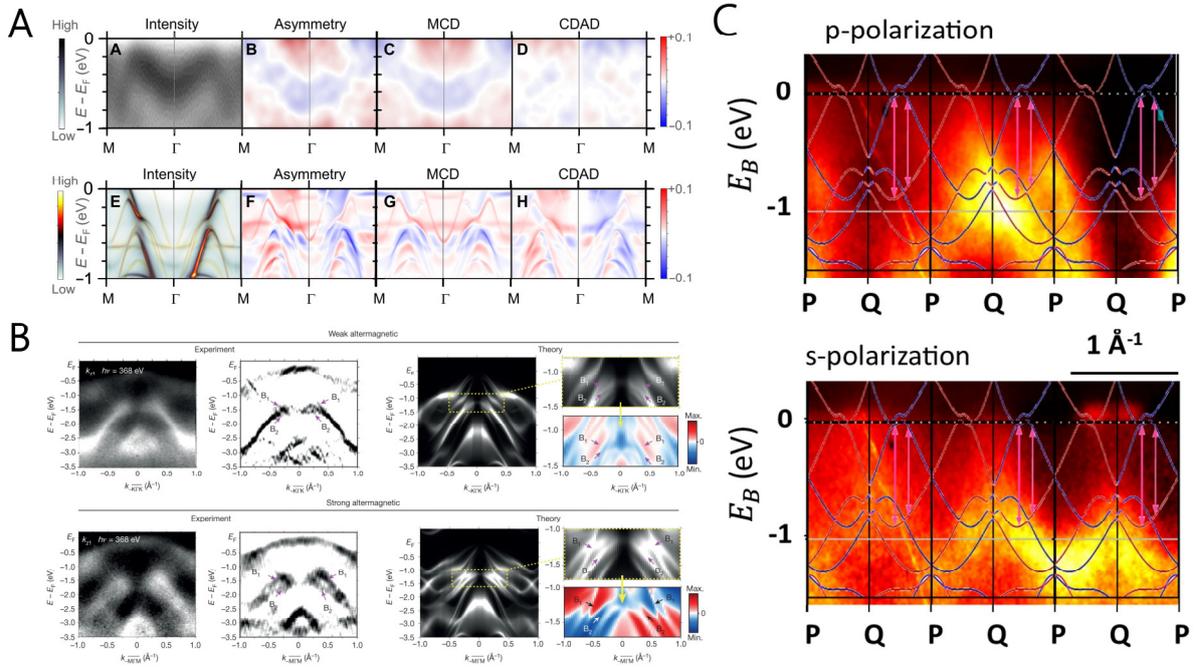

**Figure 3**. A) The upper and lower panels show the experimental measurements and theoretical calculations of the intensity map, intensity asymmetry, magnetic circular dichroism, and circular dichroism in the angular distribution for the $RuO_2$ crystal, respectively. Reproduced with permission.[59] Copyright 2024, American Association for the Advancement of Science. B) The upper and lower panels show the soft X-ray ARPES band maps and one-step ARPES simulations for the MnTe crystal, measured along the Γ-K and Γ-M paths, respectively. Reproduced with permission.[63] Copyright 2024, Springer Nature. C) Band splitting near the Fermi energy of CrSb crystal, with spin-integrated SX-ARPES intensity measured along the high-symmetry P-Q path using p-polarized and s-polarized photons, overlaid with spin-resolved band structure calculations. Reproduced with permission.[66] Copyright 2024, American Association for the Advancement of Science.

## 2.2. Anomalous and Spin Transport Properties

Altermagnets have garnered widespread attention not only for their spin-splitting phenomena but also for their spontaneous breaking of $\mathcal{T}$. This symmetry violation induces ferromagnetic-like magnetic transport properties, which were previously considered absent in collinear AFMs. These effects include the AHE, ANE, ATHE, MOEs, and $\mathcal{T}$-odd spin current. Below, we will review the progress of research on these magnetic transport properties in AMs one by one.

2.2.1. AHE, ANE, and ATHE

In 1878, Hall discovered that when a current-carrying conductor is placed in an external magnetic field, the electrons are deflected to one side of the conductor due to the Lorentz force, generating a transverse voltage drop.[83] This phenomenon later became known as the Hall effect. In 1881, Hall further observed that FMs exhibit a transverse Hall current even in the absence of an external magnetic field when subjected to a longitudinal charge current.[84] This phenomenon was termed the AHE, marking an early understanding of electron transport in magnetic materials. A century later, the concept of the Berry phase was introduced by Berry, providing a quantitative method to determine the magnitude of the intrinsic anomalous Hall conductivity (AHC).[85] In this geometric framework, the derivative of the Berry phase, known as the Berry curvature, is summed over all occupied states in the Fermi sea to yield the AHC.[86,87] This discovery reignited interest in the anomalous transport properties of magnetic materials. In 2004, Yao et al. employed first-principles calculations within this geometric framework, calculating the AHC of bcc Fe, which is in good agreement with experimental measurements, thereby establishing the intrinsic origin of its AHC.[88] Beyond the intrinsic mechanism, extrinsic mechanisms, such as skew scattering and side-jump effects driven by disorder, also contribute to the AHE. Both intrinsic and extrinsic contributions can be attributed to the breaking of $\mathcal{T}$ in combination with SOC.[89–91]

From a symmetrical standpoint, the AHC ($\sigma$) is proportional to $\boldsymbol{E} \times \boldsymbol{j_H}$, where $\boldsymbol{E}$ represents the electric field, and $\boldsymbol{j_H}$ denotes the Hall current density. In systems with $\mathcal{T}$ or $\mathcal{T}S$ symmetry ($S$ is $t$ or $\mathcal{P}$), the $\boldsymbol{E}$ is even under $\mathcal{T}(\mathcal{T}t)$ and odd under $\mathcal{T}\mathcal{P}$, while $\boldsymbol{j_H}$ behaves oppositely, odd under $\mathcal{T}(\mathcal{T}t)$ and even under $\mathcal{T}\mathcal{P}$. Consequently, the AHC σ transforms as -σ under $\mathcal{T}$ or $\mathcal{T}S$, leading to the conventional wisdom that the AHE is absent in conventional collinear AFMs. The emergence of AMs challenged this conventional understanding, establishing them as promising candidates for generating large Hall currents.

Before the concept of altermagnetism was proposed, Ghimire et al. observed a large intrinsic AHE in the chiral collinear antiferromagnetic material $CoNb_3S_6$ in 2018. They attributed this large AHE to the formation of a complex magnetic texture and the presence of Weyl nodes near the Fermi level, which could not be explained by the collinear AFM structure alone.[92] In 2020, Šmejkal et al. identified the AHE mechanism in $RuO_2$ and $CoNb_3S_6$, demonstrating that it arises from the combined effects of collinear AFM order and crystal symmetries, rather than from the magnetic structure alone.[70] They coined the term "crystal Hall effect" (CHE) to distinguish it from the classical AHE (**Figure 4A**). The CHE exhibits strong anisotropy relative to the Néel vector direction due to the anisotropic crystal environment of magnetic atoms. For example, in $RuO_2$, the CHE is turned off when the Néel vector aligns with the [001] direction and when it aligns with the [110] direction. For chiral AFM $CoNb_3S_6$, the sign of CHE can be reversed by altering the structure chirality, as the two chiral states are connected by a mirror symmetry ($M_x$) that changes the sign of $\sigma_{xy}$. Following theoretical predictions, Feng et al. experimentally demonstrated the AHE in altermagnetic $RuO_2$ by reorienting the Néel vector from the [001] easy axis to [110] direction via a magnetic field (**Figure 4B**). The reported AHC exceeded 1000 $\Omega^{-1}cm^{-1}$ under a magnetic field of 50 T.[93] In addition to applying a magnetic field, element doping offers an alternative approach to manipulating the AHE in antiferromagnetic metals. For instance, doping Cr atoms into $RuO_2$ at Ru sites reorients the Néel vector from [001] to [110] axis while maintaining the altermagnetic order. As a result, Cr-doped $RuO_2$ exhibits an AHE even without an external magnetic field.[94] Recently, the nature of the altermagnetism in $RuO_2$ remains a topic of debate.[95,96] The AHE observed in the Cr-doped rutile, $Ru_{0.8}Cr_{0.2}O_2$, is considered to originate from the magnetic Cr ions.[97]

In addition to $RuO_2$ and $CoNb_3S_6$, many other AMs have demonstrated AHE, such as organic conductors κ-(ET)$_2$X,[98] perovskites,[99–101] MnTe,[102,103] $Mn_5Si_3$,[104,105] GdAlSi,[106] CrAs[107], CrSb[69] and so on.[108–110] Among them, MnTe and $Mn_5Si_3$ have experimentally shown AHE closely linked to the Néel vector (**Figures 4C-D**).[102,104] Notably, MnTe has a magnetic easy axis along the [1$\bar{1}$00] direction, with a relativistic magnetic point group of m′m′m, enabling spontaneous AHE in zero magnetic fields. Recently, Zhou et al. first realized the manipulation of altermagnetic order and AHE in CrSb films through crystal symmetry, a method distinct from previous approaches that rely on tuning the Néel vector orientation (**Figure 4E**).[69] Similar to conventional magnetic topological semimetals, AMs can exhibit significant Berry curvature at Weyl points and nodal lines, leading to topologically enhanced anomalous transport properties. For instance, Laha et al. reported a large AHC of ~1310 $\Omega^{-1}cm^{-1}$ at 2K in altermagnetic Weyl semimetal GdAlSi [106]. Similar topologically enhanced AHE has been predicted in $CaCrO_3$ and CrAs.[100,107] However, the AHE in many AMs is strongly temperature-dependent, with large AHC observed only at low temperatures, which rapidly diminish or vanish near room temperature.[102,104,106] Consequently, discovering materials that exhibit zero-field AHE induced by collinear antiferromagnetic order at room temperature is a pressing challenge. Recently, Takagi et al. reported spontaneous AHE in FeS at room temperature and zero magnetic field due to the $\mathcal{TS}$-symmetry-broken antiferromagnetic order, opening new possibilities for electrical reading and writing of antiferromagnetic information at room temperature.[111] Given the relatively small amplitude of the observed spontaneous Hall signal, identifying practical collinear AFMs with topological features, such as Weyl points, nodal lines, or nodal planes, that exhibit spontaneous AHE at room temperature and zero magnetic field is a crucial step for advancing applications of AFMs in the future.

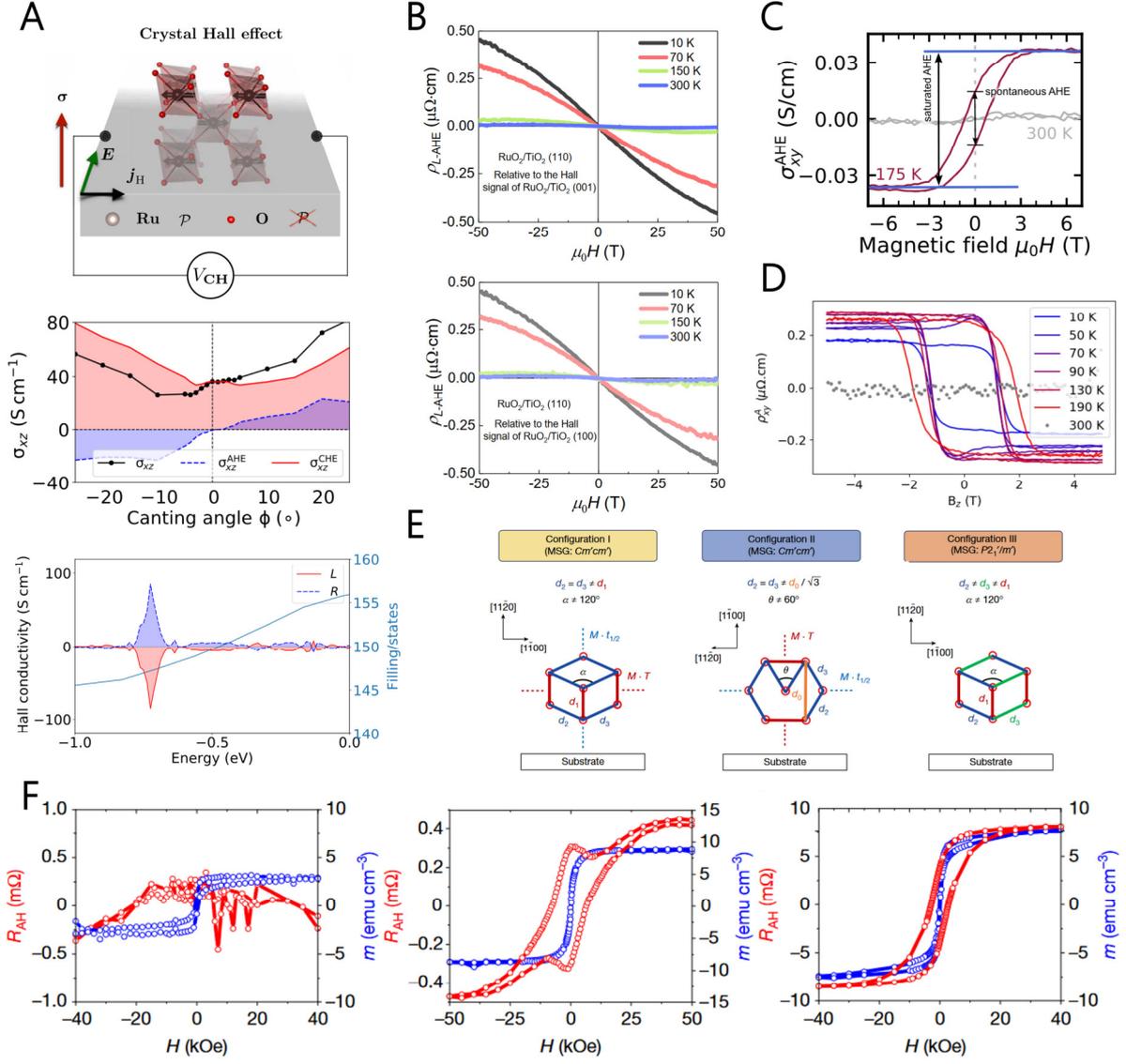

**Figure 4.** A) Crystal Hall effect generated by collinear antiferromagnetism, the dependence on the canting angle of the Hall conductivity of RuO$_2$, and the Hall conductivity of CoNb$_3$S$_6$ with left- to right-handed crystal chirality. Reproduced with permission.[70] Copyright 2020, American Association for the Advancement of Science. B) The anomalous Hall effect in RuO$_2$ crystals obtained by subtracting the Hall signals of the (001) and (100)-oriented samples, respectively. Reproduced with permission.[93] Copyright 2022, Springer Nature. C) The anomalous Hall conductivity measured in MnTe crystal at 175 K and 300 K. Reproduced under the terms of the CC-BY 4.0 license.[102] Copyright 2023, The Authors, Published by American Physical Society. D) The anomalous Hall resistivity in Mn$_5$Si$_3$ crystals as a function of magnetic field and temperature. Reproduced with permission.[104] Copyright 2024, Springer Nature. E) Crystal distortions of CrSb in configurations I, II, and III. F) Anomalous Hall resistances and out-of-plane magnetizations in configurations I, II, and III (from left to right, respectively). Reproduced with permission.[69] Copyright 2025, Springer Nature.

In addition to the longitudinal electric field ($E_x$) inducing a transverse charge current ($J_y$) (AHE),

the longitudinal temperature gradient ($-\nabla_x T$) can also generate both a transverse charge current and a transverse heat current ($J_y^Q$). These phenomena are known as the ANE[112] and ATHE[113], respectively, and are fundamental manifestations of anomalous thermal transport in spin caloritronics. These three effects are interconnected within the framework of linear response theory as follows:

$$J_\beta = \sigma_{\beta\alpha} E_\alpha - \alpha_{\beta\alpha} \nabla_\alpha T,$$

$$J_\beta^Q = T\alpha_{\beta\alpha} E_\alpha - \kappa_{\beta\alpha} \nabla_\alpha T,$$

Here, $\sigma_{\beta\alpha}$, $\alpha_{\beta\alpha}$, and $\kappa_{\beta\alpha}$ represent the AHC, anomalous Nernst conductivity (ANC), and anomalous thermal Hall conductivity (ATHC), respectively. In 2006, D. Xiao et al. demonstrated that, like the AHE, the ANE can also be described using Berry phase theory. However, unlike the AHE, where contributions arise from both the Fermi sea and the Fermi surface, the ANE predominantly stems from the region near the Fermi surface.[112] Using first-principles calculations, Xiao et al. successfully determined the ANC in the $CuCr_2Se_{4-x}Br_x$ system, with results in close agreement with experimental observations. Anomalous thermal Hall transport and anomalous charge transport are linked by the anomalous Lorenz ratio, defined as $L_{\alpha\beta} = \kappa_{\alpha\beta}/\sigma_{\alpha\beta}$. In the zero-temperature limit, this ratio approaches the Sommerfeld constant $L_0 = 2.44 \times 10^{-8} \Omega W K^{-2}$, which is famously known as the anomalous Wiedemann-Franz law. Although this relationship has been experimentally observed in FMs, less is known about its properties in collinear AFMs.[114,115]

From a symmetry perspective, the ANE and ATHE share the same symmetry requirements as the AHE. Specifically, the breaking of $\mathcal{TS}$ symmetry in AMs enables the occurrence of ANE and ATHE, both of which depend on the Néel vector and crystal symmetry. We take $RuO_2$ as an example for this review.[71] As shown in **Figure 5A**, when the Néel vector rotates within the ($\bar{1}10$) plane from the [001] axis to the [110] axis, the ANC and ATHC at 300 K exhibit strong anisotropic as functions of the polar angle. At a polar angle of 0° (with the Néel vector aligned along the [001] axis), both ANE and ATHE are suppressed due to the two glide mirror symmetries $M_{[010]}t_{1/2}$ and $M_{[100]}t_{1/2}$ in the magnetic space group $P4_2'/mnm'$. As the polar angle shifts from 0° to 90°, the $zx$ and $yz$ components, constrained by the $\mathcal{T}C_{2x\bar{y}}$ symmetry, are nearly identical and increase progressively. This is mainly attributed to the enhanced Berry curvature of three types of interband transitions in $RuO_2$ (**Figure 5B**): (1) Weyl pseudo-nodal lines within the same spin bands, (2) altermagnetic pseudo-nodal surfaces formed between opposite spin bands, and (3) altermagnetic ladder transitions between weakly spin-split bands with similar dispersion crossing the Fermi surface. The latter two transitions, characterized by unique altermagnetic features, arise from the (quasi)degeneracy of opposite spin states driven by crystal symmetry, which facilitates the spin-flip transitions. Within the Fermi energy range of $-0.2\sim0.0$ eV, the ANE is entirely dominated by spin-flip processes. Additionally, as shown in **Figure 5C**, both the ANC and ATHC exhibit significant magnitudes, reaching maxima of $-0.35$ AK$^{-1}$m$^{-1}$ and $5.5\times10^{-2}$ WK$^{-1}$m$^{-1}$ at room-temperature, respectively. Notably, due to the presence of Weyl nodes in RuO$_2$, the Wiedemann-Franz law remains valid within the nodal line energy range (−0.2∼0.2 eV) and up to temperatures of 150 K (**Figure 5D**), challenging the conventional finite-temperature violations observed in FMs.

The ANE in AMs was first experimentally observed in Mn$_5$Si$_3$.[116,117] As shown in **Figures 5E-F**, the Nernst signal $S_{yx}$ of Mn$_5$Si$_3$ exhibits clear hysteresis and saturation behavior as a function of the applied magnetic field, consistent with the field dependence of the transverse conductivity. The spontaneous ANC $\alpha_{yx}$ varies from 0.11±0.08 AK$^{-1}$m$^{-1}$ at 58 K to 0.015±0.005 AK$^{-1}$m$^{-1}$ at 216 K.[116] Although relatively weak, slight Mn doping in Mn$_5$Si$_3$ (Mn$_{5.10}$Si$_{2.90}$) can shift the Fermi level, enhancing the ANC by approximately sixfold.[117] Furthermore, first-principles calculations indicate that the ANC of CrSb remains at −0.185 AK$^{-1}$m$^{-1}$ at room temperature, with a slower decay rate at higher temperatures, demonstrating its potential for thermoelectric applications.[118] In contrast, research on the ATHE in AMs remains limited, and no experimental observations have been reported to date. Recently, Hoyer et al. discovered that the ATHE can emerge in insulating AMs and proposed candidate materials for experimental realization.[119] These studies mentioned above systematically uncover the origins of the AHE, ANE, and ATHE in AMs from both theoretical and experimental perspectives, while proposing various effective regulation methods. They highlight that AMs possess exceptional anomalous transport properties, positioning them as key players in the future advancements of spintronics and spin caloritronics.

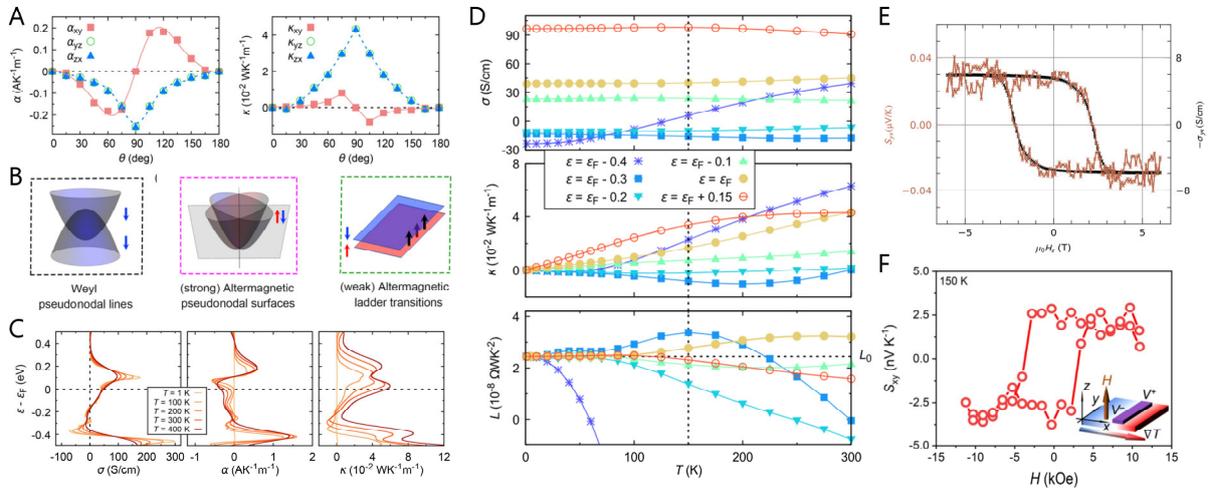

**Figure 5.** A) At T = 300 K, the ANC and ATHC of RuO$_2$ as functions of the polar angle, which describes the rotation of the Néél vector within the ($\bar{1}$10) plane. B) Three distinct types of interband transitions in RuO$_2$: Weyl pseudonodal lines, altermagnetic pseudonodal surfaces, and altermagnetic ladder transitions, respectively. C) AHC, ANC, and ATHC as a function of Fermi energy for different temperatures. D) Temperature dependence of AHC, ATHC, and anomalous Lorenz ratio with different Fermi energies. Reproduced under the terms of the CC-BY 4.0 license.[71] Copyright 2024, The Authors, Published by American Physical Society. E) Magnetic field dependence of transverse Nernst signal for Mn$_5$Si$_3$ at 216 K. Reproduced under the terms of the CC-BY 4.0 license. [116] Copyright 2024, The Authors, Published by arXiv. F)

### 2.2.2 MOEs

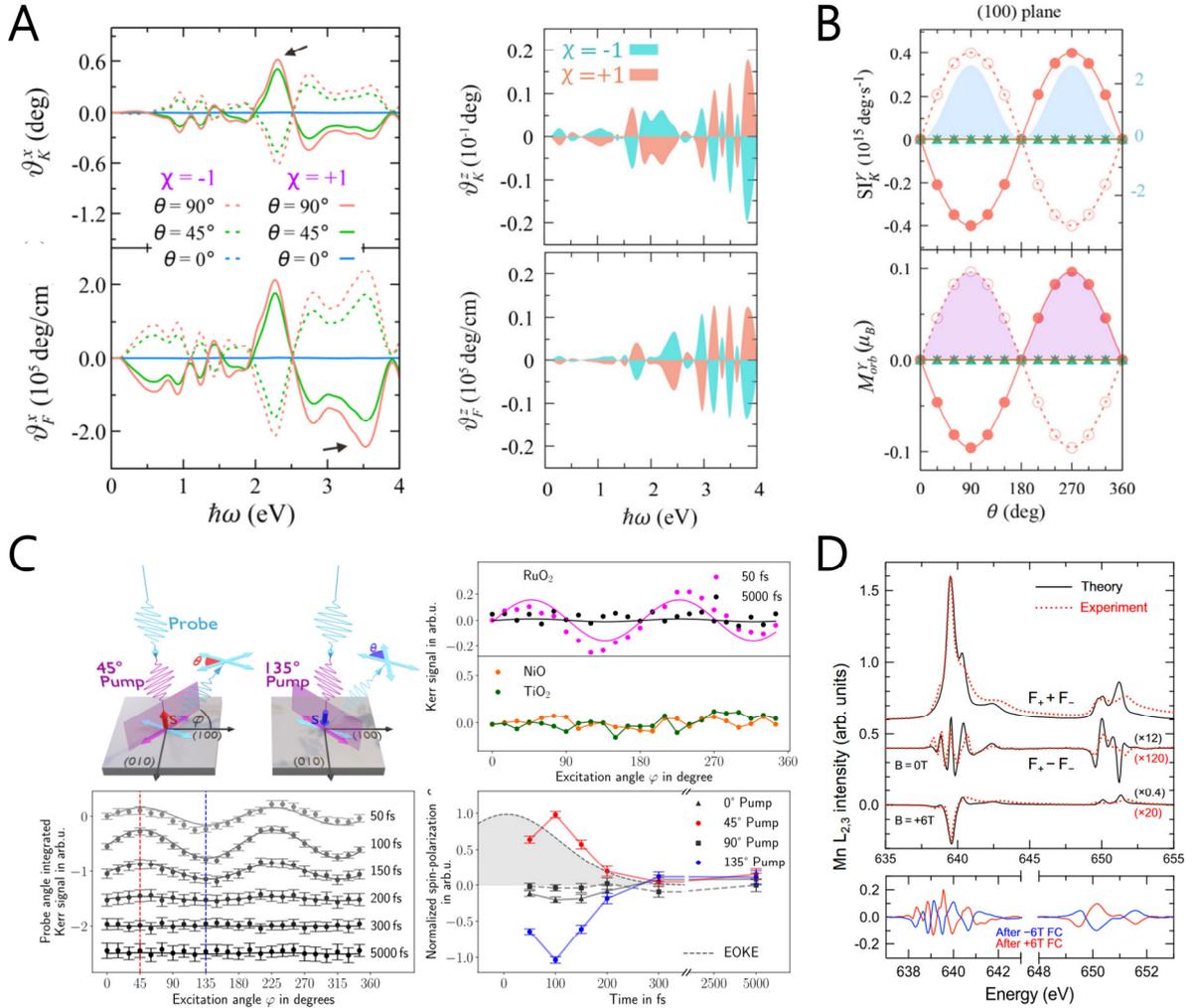

**Figure 6.** A) Kerr and Faraday rotation angles of the $RuO_2$ (left) and $CoNb_3S_6$ (right) crystal. B) The Kerr rotation angle, magneto-crystalline anisotropy energy, and orbital magnetic moment for the two chiral states of $RuO_2$ with the Néel vector rotating within the (100) plane. Reproduced under the terms of the CC-BY 4.0 license.[124] Copyright 2021, The Authors, Published by American Physical Society. C) Sketch of the time-resolved polar magneto-optical Kerr effect setup and the corresponding magneto-optical signals for $RuO_2$ crystal at different excitation angles and times. Reproduced under the terms of the CC-BY 4.0 license.[125] Copyright 2024, The Authors, Published by arXiv. D) Theoretical (black) and experimental (red dotted) Mn $L_{2,3}$-edge spectra for light propagating along the c axis. Reproduced under the terms of the CC-BY 4.0 license.[126] Copyright 2024, The Authors, Published by American Physical Society.

MOEs are among the oldest physical phenomena in magnetic materials, reflecting the fundamental interactions between light and magnetism. In 1846, Faraday discovered that when linearly polarized light passes through a glass plate placed in an external magnetic field, the plane of polarization of the transmitted light is rotated.[120] Thirty years later, Kerr observed a similar rotation effect in reflected light.[121] These phenomena were later named the magneto-optical Faraday effect (MOFE) and the magneto-optical Kerr effect (MOKE), respectively. MOEs can be regarded as the ac (finite frequency) counterpart of the AHE, sharing a similar physical origin and symmetry requirements, namely the breaking of $\mathcal{TS}$ symmetry. As a result, MOEs can also occur in AMs. Following the discovery of the crystal Hall effect in $RuO_2$ by Libor Šmejkal and his colleagues, Samanta et al. identified the crystal MOEs in $SrRuO_3$ thin films with collinear compensated antiferromagnetic order.[122] Mazin et al predicted unconventional magnetism and sizable MOKE in $FeSb_2$ alloys.[123] Furthermore, Zhou et al. introduced the concept of the crystal chiral MOE, demonstrating that its sign is governed by both the crystal chirality and the direction of the Néel vector, using $RuO_2$ and $CoNb_3S_6$ as examples (**Figure 6A**).[124] This distinguishes it from conventional MOEs, which are controlled solely by the Néel vector direction in noncollinear AFMs and FMs. The magnitudes of MOEs strongly depend on the Néel vector direction. The maximum Kerr and Faraday rotation angles in $RuO_2$ are $0.62°$ and $2.4 \times 10^5$ deg/cm, respectively, comparable to conventional FMs, highlighting its significant potential for future magneto-optical devices. They also found that both the MOEs and magneto-crystalline anisotropy energy are positively correlated with the orbital magnetic moments, a relationship identified in AFMs for the first time (**Figure 6B**). Furthermore, Sun et al. proposed a novel type-III multiferroic material, bilayer $MnPSe_3$, where its sliding ferroelectricity could serve as a switch for reversing both the altermagnetic spin orientation and the sign of the MOKE.[55]

Experimentally, Iguchi et al. measured the MOE in the organic AM κ-(BEDT−TTF)$_2$Cu[N(CN)$_2$]Cl. The MOKE emerged near the Néel temperature and exhibited a nonlinear field dependence in the antiferromagnetic phase.[127] Weber et al. performed magneto-optical experiments on $RuO_2$ films, revealing that the *d*-wave characteristic of the spin-split energy bands in $RuO_2$ facilitates the selective excitation of hot carriers in spin-resolved states using linearly polarized light pulses (**Figure 6C**). This process induces a transient spin response, enabling femtosecond-scale control over both the sign and magnitude of the photoinduced spin polarization, depending on the alignment of the light polarization vector with respect to the two oppositely staggered spin subsystems of the AMs.[125] Similarly, Gray et al. utilized pump-probe magneto-optical techniques in the Kerr configuration to study ultrafast spin dynamics in epitaxial MnTe(001)/InP(111) thin films. They detected optical phonon modes at frequencies of 3.6 THz and 4.2 THz, as well as a 55 GHz oscillation modulated by the magnetic field, which persisted even above the Néel temperature ($T_N$=310 K).[128] Furthermore, Hariki et al. investigated the x-ray magnetic circular dichroism (XMCD) in α-MnTe,[126] a type of MOE that shares the same requirements as the MOKE and MOFE. As shown in **Figure 6D**, their computational results are in excellent agreement with the experimentally measured XMCD spectra. This study elucidates the presence of a Néel vector lying in the plane perpendicular to the light propagation vector in α-MnTe. These results highlight the effectiveness of magneto-optical tools in investigating altermagnetism and underscore the utility of MOEs in uncovering the physical properties of altermagnets.

2.2.3. Spin splitter torque

The generation of spin currents is a fundamental aspect of spintronics, arising primarily through two distinct mechanisms: ferromagnetic exchange splitting in FMs and the spin Hall effect (SHE) or spin Nernst effect (SNE) in nonmagnetic systems. Ferromagnetic exchange splitting produces $\mathcal{T}$-odd longitudinal spin-polarized currents,[129] while SHE and SNE generate $\mathcal{T}$-even transverse spin currents, induced by a longitudinal charge current or temperature gradient.[130,131] These two types of spin currents enable distinct functionalities in spintronic applications: longitudinal $\mathcal{T}$-odd currents drive spin-transfer torque (STT) in magnetic random access memories,[132,133] whereas transverse $\mathcal{T}$-even currents power spin-orbit torque (SOT) devices.[134,135] In addition to exhibiting FMs-like anomalous transport properties induced by symmetry breaking, AMs also possess a $\mathcal{T}$-odd spin current via spin splitter torque (SST), which arises directly from the spin split electronic bands, as depicted in **Figure 7A**.[136]. Such a phenomenon is absent in conventional AFMs that are constrained by the $\mathcal{TS}$ symmetry. Unlike traditional mechanisms such as SOT, which relies on SOC, or STT, which depends on ferromagnetic exchange interactions, the SST is rooted in nonrelativistic collinear antiferromagnetic order, requiring neither SOC nor ferromagnetic structures. This fundamentally circumvents the short spin diffusion lengths caused by relativistic effects in SOT and the dependency on FMs in STT. The generated spin current exhibits $\mathcal{T}$-odd characteristics (shared with STT) and enables a pure spin current coexisting with zero parallel charge current. Taking $RuO_2$ as an example, its spin-splitting angle is primarily governed by the anisotropy of the spin sublattice, with a negligible influence from SOC. The calculated charge-spin conversion ratio, $\sigma_{bc}^{odd,a}/\sigma \approx 28\%$, corresponds to an angle of 34° between the spin-up and spin-down transport channels, which is significantly larger than that of conventional materials such as Pt and β-W (**Figure 7A-C**).[134] The spin currents driven by SHE or SNE are also reported in other AMs, such as κ-(BEDT−TTF)$_2$Cu[N(CN)$_2$]Cl,[137] $V_2Se_2O$,[76] $V_2Te_2O$,[138] $Cr_2Te(Se)_2O$,[139] $GdFeO_3$-type perovskite,[140] and twisted bilayer VOBr.[81]

The $\mathcal{T}$-odd spin current in $RuO_2$ has been successfully measured in experiments through spin-torque ferromagnetic resonance (ST-FMR) measurements, yielding results consistent with theoretical predictions.[141–143] As shown in **Figures 7D-E**, in the (001)-oriented sample, no unconventional SOT was observed, and the signal aligns with the conventional SHE model. In contrast, for the (101)-oriented sample, when the Néel vector is tilted out of the sample plane, a significant unconventional SOT was detected under an electric field applied along the [010] direction, characterized by an additional damping-like component, confirming the presence of a nonrelativistic spin-splitter effect (SSE).[141] At the same time, both Bai et al. and Karube et al. constructed ST-FMR measurement setups and confirmed the existence of SST in $RuO_2$. As shown in **Figure 7F**, Bai et al. found that the spin torque efficiency of the $RuO_2$ (100)/Py sample is significantly higher than that of $RuO_2$ (110)/Py, indicating the presence of crystal orientation-dependent SST. They also found that the generated $\mathcal{T}$-odd spin current is independent of spin-orbit coupling, and its polarization direction is modulated by the Néel vector, opening up new possibilities for achieving controllable spin currents.[142] Karube et al. decomposed the ST-FMR raw data into symmetric (Lorentzian) and antisymmetric (derivative) components, corresponding to the damping-like torque and field-like torque. Their analysis revealed that in the $RuO_2$ (101) crystal plane, spin polarization includes not only the conventional SHE-induced *y*-component but also unconventional *x*- and *z*-components, confirming the contribution of the SSE (**Figure 7G**).[143] Based on this altermagnetic SSE, $RuO_2$ can achieve efficient spin-to-charge conversion,[144,145] which has also been theoretically predicted in 2D AMs, such as monolayer $V_2Te(Se)_2O$ and $Cr_2Te(Se)_2O$.[76,138,139] Furthermore, the SSE in $RuO_2$ enables field-free switching of perpendicular magnetized FMs at room temperature and serves as a key mechanism driven by a large spin Hall angle and a long spin diffusion length.[146] These properties allow the spin splitter to combine the material versatility of SOT with the long-range spin transport advantage of STT, paving a novel path for high-efficiency, highly integrated spintronic devices.

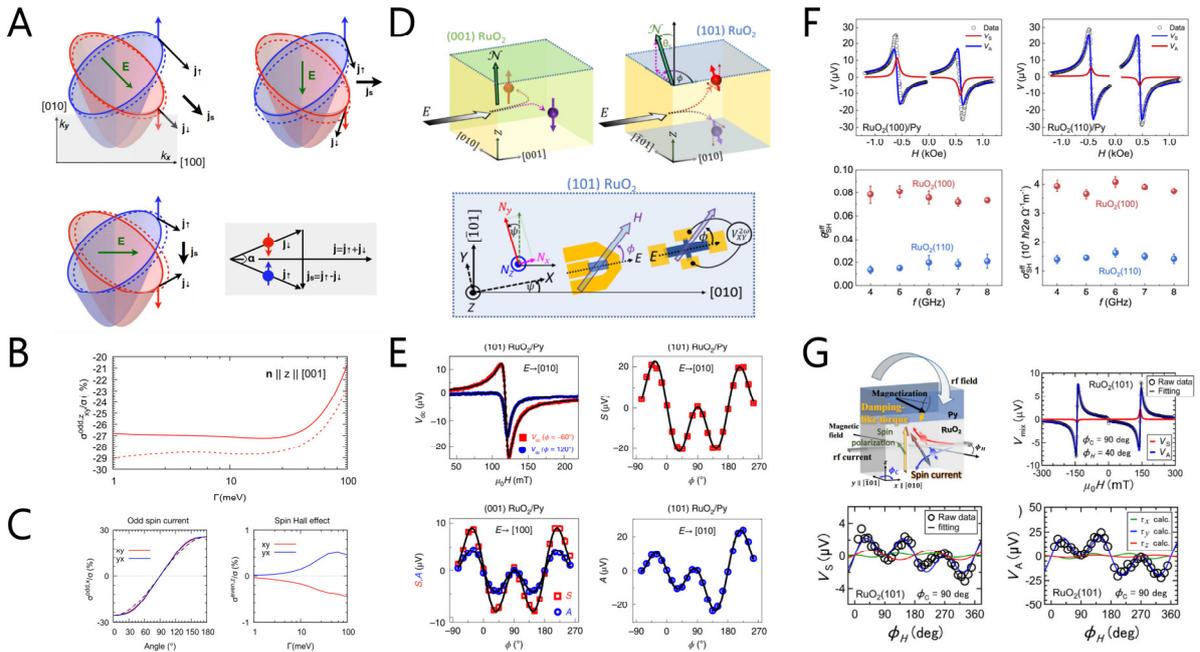

**Figure 7.** A) Schematic of band splitting and SST in altermagnetic $RuO_2$ crystal. B) The $\mathcal{T}$-odd charge-spin conversion ratio calculated with (solid line) and without (dashed line) spin-orbit coupling, with the Néel vector along the [001] direction. C) Left: $\mathcal{T}$-odd spin-splitter charge-spin conversion ratio as a function of the in-plane Néel vector angle. Right: $\mathcal{T}$-even spin-Hall charge-spin conversion ratio as a function of the scattering rate. Reproduced under the terms of the CC-BY 4.0 license.[136] Copyright 2021, The Authors, Published by American Physical Society. D) Illustration of spin current generation due to the antiferromagnetic SHE for a (001) and (101)-oriented $RuO_2$ film. Sample schematics and definition of coordinate axes for the (101)-oriented sample. E) The ST-FMR signals, symmetric and antisymmetric resonance

amplitudes as a function of in-plane magnetic field angle for (101) and (001)-oriented RuO$_2$(6 nm)/Py(4 nm) samples with electric fields applied in different directions. Reproduced with permission.[141] Copyright 2022, Springer Nature. F) Top panels: ST-FMR spectra of the RuO$_2$(100)/Py and the RuO$_2$(110)/Py samples. Bottom panels: the calculated spin torque efficiency $\theta_{SH}^{eff}$ and spin torque conductivity at different microwave frequencies in RuO$_2$(100) and RuO$_2$(110) films. Reproduced under the terms of the CC-BY 4.0 license.[142] Copyright 2022, The Authors, Published by American Physical Society. G) Top panels: Schematic image of ST-FMR measurement and applied in-plane magnetic field dependence of ST-FMR raw data. Bottom panels: Symmetric and antisymmetric voltage amplitudes depending on the applied field angle for RuO$_2$(101). Reproduced under the terms of the CC-BY 4.0 license.[143] Copyright 2022, The Authors, Published by American Physical Society.

## 3. Ncl-AFMs

Beyond the commonly studied collinear antiferromagnetic systems, there exists a wide range of systems with noncollinear spin structures. These systems can be characterized by scalar spin chirality

$$\chi = \sum_{<ijk>} \vec{S_i} \cdot (\vec{S_j} \times \vec{S_k})$$

or vector spin chirality

$$\kappa = \frac{2}{3\sqrt{3}} \sum_{<ij>} [\vec{S_i} \times \vec{S_j}]_z$$

which are used to describe noncoplanar antiferromagnetic and coplanar noncollinear antiferromagnetic structures, respectively (where $S_{i,j,k}$ represents three noncollinear spins).[147,148] Here, we refer to these as ncl-AFMs. As the name suggests, in ncl-AFMs, the angles between magnetic moments on different atoms are not 180° (as in traditional AFMs) but instead are arranged at specific angles, often forming triangular or kagome lattices. More intriguingly, these ncl-AFMs not only exhibit zero net magnetization but also possess spontaneous breaking of $\mathcal{TS}$ symmetry. As a result, they exhibit ferromagnet-like spin-polarized currents, anomalous transport and spin transport properties. Below, we take the extensively studied Mn$_3$X (X = Ga, Ge, Sn, Rh, Ir, Pt) family as an example to review the spin polarization properties and transport characteristics of ncl-AFMs.

### 3.1. Chiral spin texture and $\mathcal{T}$-odd spin current

The magnetic configuration of the Mn$_3$X family was experimentally detected as early as the last century,[149–154] but its spin-splitting phenomenon and spin-polarized current were not theoretically demonstrated until 2014 by Chen et al. and 2017 by Železný et al., respectively.[22,155] As shown in **Figure 8A**, the Mn$_3$X family can be categorized into two groups: the negative chiral state $\kappa = -1$ with a hexagonal lattice structure (space group $P6_3/mmc$, No. 194) for X = Ga, Ge, Sn, and the positive chiral state $\kappa = +1$ with a face-centered cubic lattice structure (space group $Pm\bar{3}m$, No. 221) for X = Rh, Ir, Pt. The former features a kagome-type lattice formed by Mn sublattices on the [0001] plane, which similarly emerges on the [111] plane of the latter. This unique magnetic configuration enables Mn$_3$X to lift Kramers spin degeneracy and exhibit $\mathcal{T}-$odd spin current as well as spin current angle even without considering spin-orbit coupling, as illustrated in **Figure 8B**. This behavior highlights distinct symmetry-driven mechanisms compared to conventional SHE. Ncl-AFMs also exhibit a significant spin Hall response. For example, the Mn$_3$X family demonstrates large and highly anisotropic SHE, with Mn$_3$Ga displaying a spin Hall conductivity of up to 600 $(\hbar/e)(\Omega cm)^{-1}$[156] and Mn$_3$Ir exhibiting a significant spin Hall angle (0.35) in (001)-oriented thin films.[157] In contrast to FMs, where spin-polarized currents split into two oppositely aligned spin channels, ncl-AFMs lack balanced spin currents; instead, these currents are oriented in multiple directions. Furthermore, the absence of a net magnetic moment ensures that both the integral of the spin of all electrons and the integral of spin times velocity vanish, leading to no equilibrium spin current. When an electric field is applied, electrons at the Fermi level are redistributed, generating both the longitudinal and transverse spin current in the AFM junction (**Figures 8C-D**).

To measure spin current, a SHE detector utilizing Mn$_3$Sn single crystals combined with a ferromagnetic NiFe layer was developed, successfully detecting spin accumulation within the Mn$_3$Sn crystal (**Figures 8E-F**).[158] This accumulation, characterized by $\mathcal{T}$-odd spin currents arising from momentum-dependent spin splitting induced by noncollinear magnetic ordering, is termed the magnetic SHE. The sign of this effect is reversed when the direction of the triangularly arranged magnetic moments is flipped. Specifically, due to its unique magnetic group symmetry, Mn$_3$Sn can generate an out-of-plane spin current when an in-plane charge current is applied, resulting in an out-of-plane anti-damping torque that drives field-free switching of a ferromagnet.[159] Compared to conventional $\beta$-Ta-based SHE devices, which require a 300 Oe external field and achieve only 17% magnetization switching efficiency, Mn$_3$Sn-based magnetic SHE devices achieve 60% switching efficiency without an external field (**Figure 8G**). This phenomenon has also been observed in noncollinear antiferromagnetic Mn$_3$Pt thin films with a face-centered cubic structure, which exhibits high crystalline symmetry and was initially presumed to lack anisotropy in transport properties. However, Cao et al. and Novakov et al. reported an anomalous anisotropy of spin current in the (001) film of Mn$_3$Pt, arising from both $\mathcal{T}$-odd and $\mathcal{T}$-even SHE (**Figure 8H**).[160,161]. Recently, Go et al. further discovered that the asymmetric local crystal potential in Mn$_3$Sn can generate a noncollinear spin current, which can be used to switch the chiral spin texture of Mn$_3$Sn even in the absence of an external magnetic field.[162] These novel phenomena significantly broaden the potential applications of ncl-AFMs.

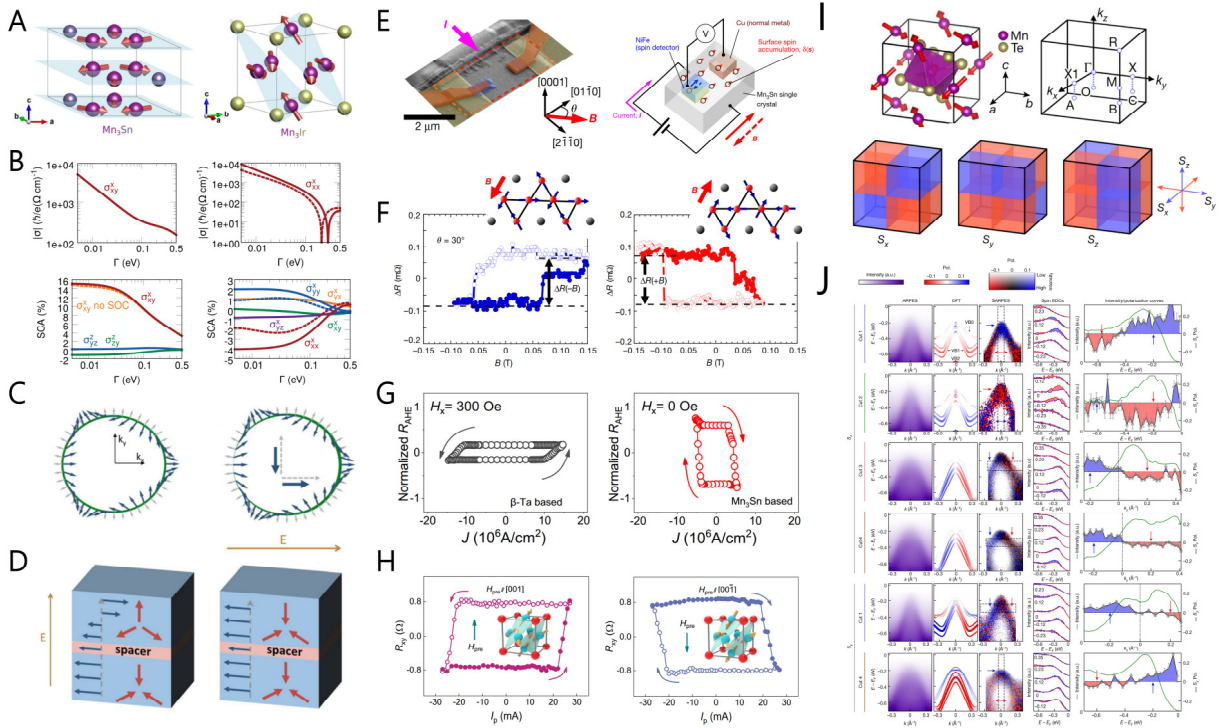

**Figure 8.** A) Crystal and magnetic structures of Mn$_3$Sn, Mn$_3$Ga, Mn$_3$Ge (left panels) and right: Mn$_3$Ir, Mn$_3$Rh, Mn$_3$Pt (right panels). B) The magnitude of the spin current (top) and spin current angle (bottom) for Mn$_3$Sn (left) and Mn$_3$Ir (right). C) The simplified Fermi level of a noncollinear antiferromagnet with and without an applied electric field, respectively. The green line represents the Fermi level, while the blue and gray arrows denote the mean values of spin and velocity, respectively. D) Parallel (left) and antiparallel (right) states of the AFM junction. Reproduced under the terms of the CC-BY 4.0 license.[22] Copyright 2017, The Authors, Published by American Physical Society. E) Scanning electron microscope image of the spin-accumulation device and schematic diagrams of the measurement setup. F) The magnetic field dependence of the resistance between the NiFe and Cu electrodes at room temperature. Reproduced with permission.[158] Copyright 2019, Springer Nature. G) The current-induced magnetization switching curve of conventional SHE-based SOT device at the external magnetic field of 300 Oe and novel magnetic SHE Mn$_3$Sn-based device with the absence of external magnetic field. Reproduced with permission.[159] Copyright 2022, Springer Nature. H) The switching polarity of the magnetic SHE in Mn$_3$Pt/Ti/CoFeB/MgO/SiO$_2$ with premagnetization fields along opposite directions. Reproduced with permission.[161] Copyright 2023, Springer Nature. I) Crystal structure, the first BZ of MnTe$_2$, and the schematics for the sign of the $S_x$, $S_y$, and $S_z$ polarization in the 3D BZ. J) The spin-integrated ARPES band dispersion, DFT-calculated bulk bands, SARPES band dispersion, spin-resolved EDCs, spin-integrated ARPES intensity (green curve), and spin polarization (symbols and solid black curve) versus binding energy or momentum (from left to right) in different regions (from top to bottom) of the reciprocal space of MnTe$_2$. Reproduced with permission.[163] Copyright 2024, Springer Nature.

Recent observations of large finite tunneling magnetoresistance at cryogenic temperatures in single-sided antiferromagnetic tunnel junctions, using noncollinear antiferromagnetic Mn₃Sn and ferromagnetic CoFeB as electrodes, confirm significant spin polarization in Mn₃Sn.[164] This observation aligns with theoretical predictions of lifted Kramers spin degeneracy caused by symmetry breaking in ncl-AFMs. Importantly, the spin-splitting electronic band of ncl-AFMs had been directly observed in MnTe₂ crystal by spin-resolved and angle-resolved photoemission spectroscopy measurements.[163] As shown in **Figure 8I**, the crystal space group, magnetic space group, and spin space group of ncl-AFM MnTe₂ are all Pa-3, with broken $\mathcal{PT}$ and $U\tau_{1/2}$ symmetries (unitary spin inversion followed by fractional translation). The four Mn sublattices with distinct spin orientations in MnTe₂ are connected by mirror symmetries $M_x$, $M_y$, and $M_z$. These symmetries result in a unique plaid-like spin texture with alternating polarizations on each side of a high-symmetry plane. For instance, $M_z$ symmetry prohibits in-plane spin components on the high-symmetry $k_z=0$ plane. However, in-plane spin splitting appears along the non-high-symmetry O-A-B-C plane even without SOC. As depicted in **Figure 8J,** the spin-resolved and angle-resolved photoemission spectroscopy measurements successfully revealed spin-polarized electronic band structures on the O-A-B-C plane in MnTe₂, providing direct evidence for the breaking of $\mathcal{T}S$ symmetry induced by noncollinear magnetic ordering. Consequently, a wide range of anomalous transport properties may also be observed in ncl-AFMs, as discussed in the following section.

**3.2. Anomalous and Spin Transport Properties**
3.2.1 AHE, ANE and ATHE
The AHE in coplanar ncl-AFMs was first discovered in Mn₃Ir crystal in 2014 through symmetry analysis and first-principles calculations. The intrinsic AHC was found to reach as high as 218 $\Omega^{-1}cm^{-1}$,[155] a value comparable to that of certain ferromagnetic transition metals.[88] This large AHE can be attributed to the absence of (111) mirror symmetry in Mn₃Ir and the strong SOC of the Ir atoms (**Figure 9A**). Beyond Mn₃Ir, other members of the Mn₃X family have also been predicted to exhibit the AHE.[156,165] Particularly, Mn₃Sn and Mn₃Ge host Weyl points near the Fermi level,[166,167] which open a gap under SOC, thereby generating large Berry curvature and AHC. The first experimental detection of the zero-field AHE in ncl-AFMs was reported in the bulk Mn₃Sn single crystal, which exhibits a considerable AHC of approximately 20 $\Omega^{-1}cm^{-1}$ at 300 K and exceeding 100 $\Omega^{-1}cm^{-1}$ at 100K (**Figure 9B**).[168] Due to the extremely weak in-plane ferromagnetism in Mn₃Sn (4-7 $m\mu_B$ per formula unit), the sign of its AHE can be reversed with a very small magnetic field (**Figure 9C**). Futhermore, the observation that the zero-field AHE significantly exceeds the magnetization-linear component of AHE demonstrates that the large AHE in Mn₃Sn originate uniquely its noncollinear antiferromagnetic order (**Figure 9D**). Later, the nearly vanishing net magnetic moment in Mn₃Ge was experimentally found to coexist with a substantial AHE originating from nonzero Berry curvature, with remarkable conductivity values of 500 $\Omega^{-1}cm^{-1}$ at 2 K and 50 $\Omega^{-1}cm^{-1}$ at room temperature (**Figure 9E**).[169] These theoretical and experimental explorations of the AHE in Mn₃Sn and Mn₃Ge have opened a new chapter in the study of ncl-AFMs. Subsequent studies have revealed significant AHE responses in diverse ncl-AFM systems, including $K_{0.5}RhO_2$,[170] γ-FeMn,[171,172] $Mn_3XN$(X=Ni, Zn, Ag, Sn, Ga),[173–182] $CoX_3S_6$(X=Ta, Nb)[183] and HoAgGe.[184] The abundant ncl-AFMs demonstrate exceptional potential as functional platforms for revolutionizing next-generation spintronic technologies.

In pursuit of higher integration density in spintronic devices, research on thin films of ncl-AFMs has become a key area of focus. In 2018, Ikeda et al. and Higo et al. reported significant AHEs in $Mn_3Sn$ thin films at room temperature. The films synthesized by Ikeda et al. contained a small amount of coexisting $Mn_2Sn$ phase, resulting in weak ferromagnetic properties at room temperature and an AHC of approximately -18.5 $Ω^{-1}cm^{-1}$ comparable to that of single crystals.[185] Higo et al., by increasing the Mn content, produced high-quality $Mn_3Sn$ films with magnetic states identical to the bulk phase (with negligible magnetization). Their measurements showed that the magnitude of the AHE was about half that of single crystals, while the coercivity was approximately 0.6 T, much greater than that of the bulk.[186] This enhanced coercivity improves the film's stability against external magnetic fields but also increases energy loss during current-driven manipulation.

Importantly, the noncollinear antiferromagnetic order and the induced AHE can be effectively tuned by external fields, such as epitaxial strain, electric field, and magnetic field, leading to intriguing new properties.[182,187–197] For example, the finite piezomagnetic coupling has been observed in $Mn_3Sn$ at room temperature due to the broken $\mathcal{T}$ symmetry, such that a strain as small as approximately -0.075±0.02% (90±24 MPa) can reverse the sign of the AHE.[196] Notably, this strain is insufficient to cause substantial changes in the electronic band structure, indicating that the sign reversal originates from its effect on the antiferromagnetic order parameters. In the $Mn_3Pt/BaTiO_3$ heterostructure, the epitaxial strain could alter the AHC of $Mn_3Pt$ thin films by more than an order of magnitude.[190] Additionally, the AHE can be switched on and off near ~360 K with a small electric field, driven by piezoelectric strain transfer from the $BaTiO_3$ substrate. A perpendicular magnetic field to the Kagome plane can also induce changes in the antiferromagnetic order of $Mn_3Sn$. This effect can lift the degeneracy of Weyl nodes, generating a large Berry curvature and an AHC that is perfectly linear with the magnetic field strength.[197] Besides, the AHE in ncl-AFMs is also closely related to factors such as sample size, film thickness, and hydrostatic pressure.[198–201]

It is important to note that the anomalous transports in coplanar ncl-AFMs only emerge when SOC is considered. However, the AHE can also occur even without SOC in noncoplanar ncl-AFMs. In these materials, electrons moving through a noncoplanar spin texture acquire a finite scalar-spin chirality, which behaves like a fictitious magnetic field, responsible for the topological Hall effect (THE).[202–211] In 2001, Shindou et al. predicted the emergence of AHE without SOC in distorted fcc lattice with noncoplanar antiferromagnetic ordered.[147] The concept of THE was subsequently proposed by Bruno et al.[203] To day, the THE has been observed in many materials,[211] such as $Nd_2Mo_2O_7$,[202] $Pr_2Ir_2O_7$,[205] $Mn_5Si_3$,[207],

KFe$_3$(OH)$_6$(SO$_4$)$_2$,[212] MnSi,[213] Mn-doped Bi$_2$Te$_3$,[214] Mn$_{1-x}$Fe$_x$Si,[208] Pt/Cr$_2$O$_3$,[215] Gd$_2$PdSi$_3$,[216] Cr$_x$(Bi$_{1-y}$Sb$_y$)$_{2-x}$Te$_3$/(Bi$_{1-y}$Sb$_y$)$_2$Te$_3$,[217] (Ca,Ce)MnO$_3$[218] and so on.

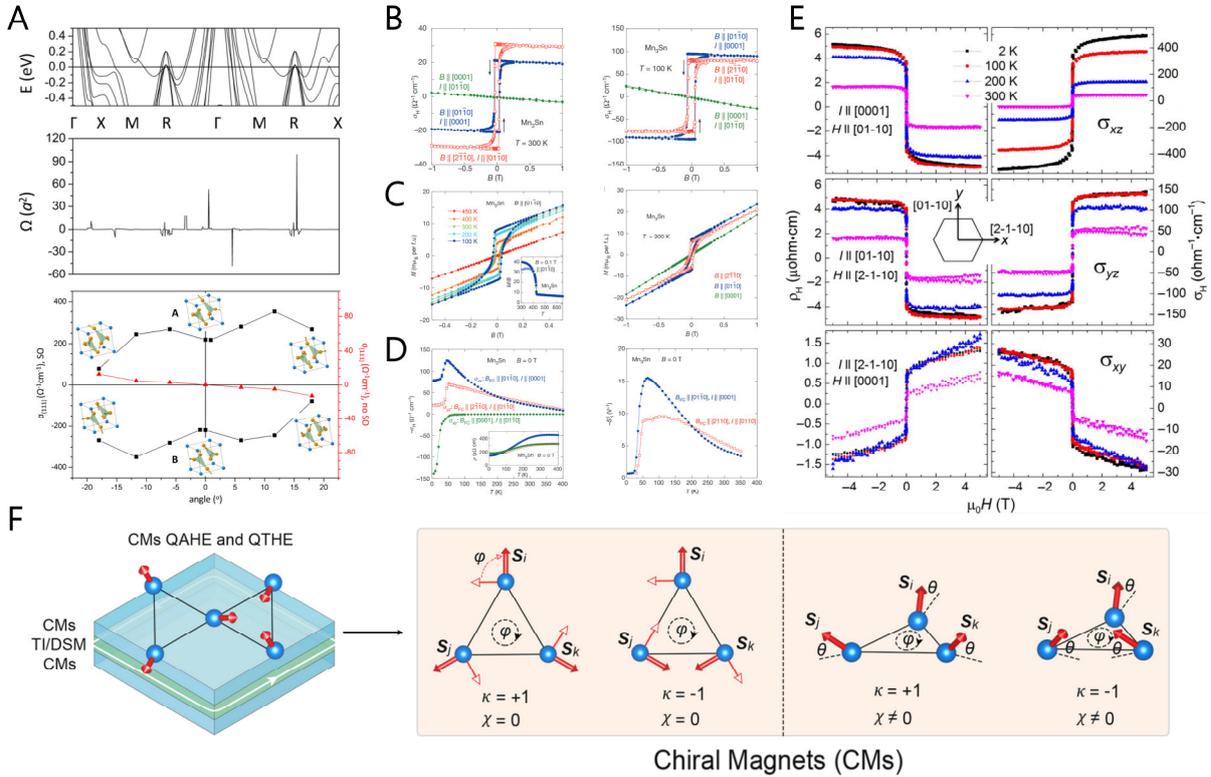

**Figure 9.** A) From top to bottom: electronic band structure and Berry curvature of Mn$_3$Ir, and the dependence of anomalous Hall conductivities of A and B phases on the Mn moment tilt angle. Reproduced under the terms of the CC-BY 4.0 license.[155] Copyright 2014, The Authors, Published by American Physical Society. B) Field dependence of the Hall conductivity measured along $B\|[2\bar{1}\bar{1}0]$, $B\|[01\bar{1}0]$, and $B\|[0001]$ at 300 K and 100 K. C) Left: Magnetization as a function of magnetic field $B\|[01\bar{1}0]$ at different temperatures. Right: Magnetization anisotropy at 300 K for $B$ along $[2\bar{1}\bar{1}0]$, $[01\bar{1}0]$, and $[0001]$. D) Left: At zero-field, AHC as a function of different temperatures. Right: Temperature dependence of $S_H^0 = -\sigma_H(B=0)/M(B=0)$ after field cooling, where $M(B=0)$ is the remanent magnetization. Reproduced with permission.[168] Copyright 2015, Springer Nature. E) The Hall resistivity (left panels) and Hall conductivity (right panels) of Mn$_3$Ge as a function of a magnetic field, measured over a temperature range of 2 K to 300 K, for different electric field and magnetic field directions. Reproduced with permission.[169] Copyright 2016, American Association for the Advancement of Science. F) Schematic illustration of Quantum AHE and Quantum THE in chiral magnets. Reproduced with permission.[219] Copyright 2016, American Chemical Society

When a nontrial band gap emerges at the Fermi level, a quantized version of AHE can be expected in both coplanar ncl-AFMs[173,219–222] and noncoplanar ncl-AFMs.[170,219,223] For example, Zhou el al. first predicted a large band gap quantized THE in noncoplanar ncl-AFMs K$_{0.5}$RhO$_2$.[170] Liu et al. predicted the quantized AHE and quantized SHE in noncoplanar ncl-AFMs NiTl$_2$S$_4$.[223] Zhou et al. demonstrate the emergence of spin-chirality-driven quantized AHE and quantized THE in kagome chiral AFMs by model analysis and first-principles calculations (**Figure 9F**),[173,219] where the quantized AHE emerges even without magnetization or SOC. Moreover, the Chern number can be periodically regulated by the interplay between vector- and scalar-spin chiralities.

In addition to AHE, ncl-AFMs also exhibit other anomalous transport phenomena, such as the ANE and ATHE. These thermal transport effects offer new perspectives on the coupling between spin current and thermodynamics in spin caloritronics. The study of the ANE and ATHE in ncl-AFMs also began with the Mn$_3$X family.[224–226] Among them, the anomalous thermal transport signals in Mn$_3$Sn crystal were the first to be experimentally detected.[225–227] As shown in Figure 10A, the transverse Nernst and thermal Hall signals in Mn$_3$Sn exhibit a step-like response to an external magnetic field, indicating that its thermal transports are dominated by Berry curvature. Additionally, the anomalous thermal and electrical conductivity closely follow the Wiedemann-Franz law across a wide temperature range (up to 400 K), suggesting that the AHE in Mn$_3$Sn is intrinsic, excluding contributions from inelastic scattering, such as phonons and magnon excitations.[225] At the same time, Ikhlas et al. discovered a large zero-field ANE in Mn$_3$Sn, which arises from a significantly enhanced Berry curvature associated with Weyl points near the Fermi energy (Figure 10B).[226] Furthermore, the large ANE observed in Mn$_3$Sn does not follow the proportional relationship between ANE and magnetization found in conventional FMs, highlighting that the ANE in noncollinear AFMs is solely determined by the Berry curvature near the Fermi level.

Similar to Mn$_3$Sn, the ANE in Mn$_3$Ge crystals and thin films were also subsequently reported.[228–230] Xu et al. systematically verified the anomalous transverse Wiedemann-Franz law in Mn$_3$Ge, finding that it is valid up to 100 K. They demonstrated that the finite-temperature violation of the Wiedemann-Franz law is caused by a mismatch between the thermal and electrical summations of Berry curvature, not by inelastic scattering (Figure 10C).[229] However, the measured ANC in Mn$_3$Sn and Mn$_3$Ge is not significant (0.2–0.4 A K$^{-1}$m$^{-1}$), which is an order of magnitude smaller than the maximum ANC (4 A K$^{-1}$m$^{-1}$) experimentally observed in the ferromagnetic topological semimetal Co$_2$MnGa.[231] Therefore, there is a need to search for materials with large ANC at room temperature. Recently, an ANC as high as 10 A K$^{-1}$m$^{-1}$ was experimentally observed in the YbMnBi$_2$ crystal with noncollinear antiferromagnetic order (a spin-canted collinear antiferromagnet).[232] This substantial ANE arises from the band topology and the possible extrinsic mechanism caused by strong SOC. YbMnBi$_2$ simultaneously exhibits small magnetization, strong anisotropy, and remarkable ANE across a broad temperature range (Figure 10D), underscoring its outstanding thermoelectric performance and potential for practical applications. Additionally, the system of ncl-AFM Mn$_3$NiN was predicted to exhibit an ANC of 1.8 A K$^{-1}$m$^{-1}$ at 200 K through first-principles calculations, significantly higher than that of conventional FMs (0–1 A K$^{-1}$m$^{-1}$).[233] The subsequent experiments also observed the ANE in the (001)-oriented Mn$_3$NiN and Mn$_3$SnN thin films.[234–236] Overall, the anomalous transport in ncl-AFMs not only reveals the unique physical properties but also offers valuable insights for the design of future spintronic devices.

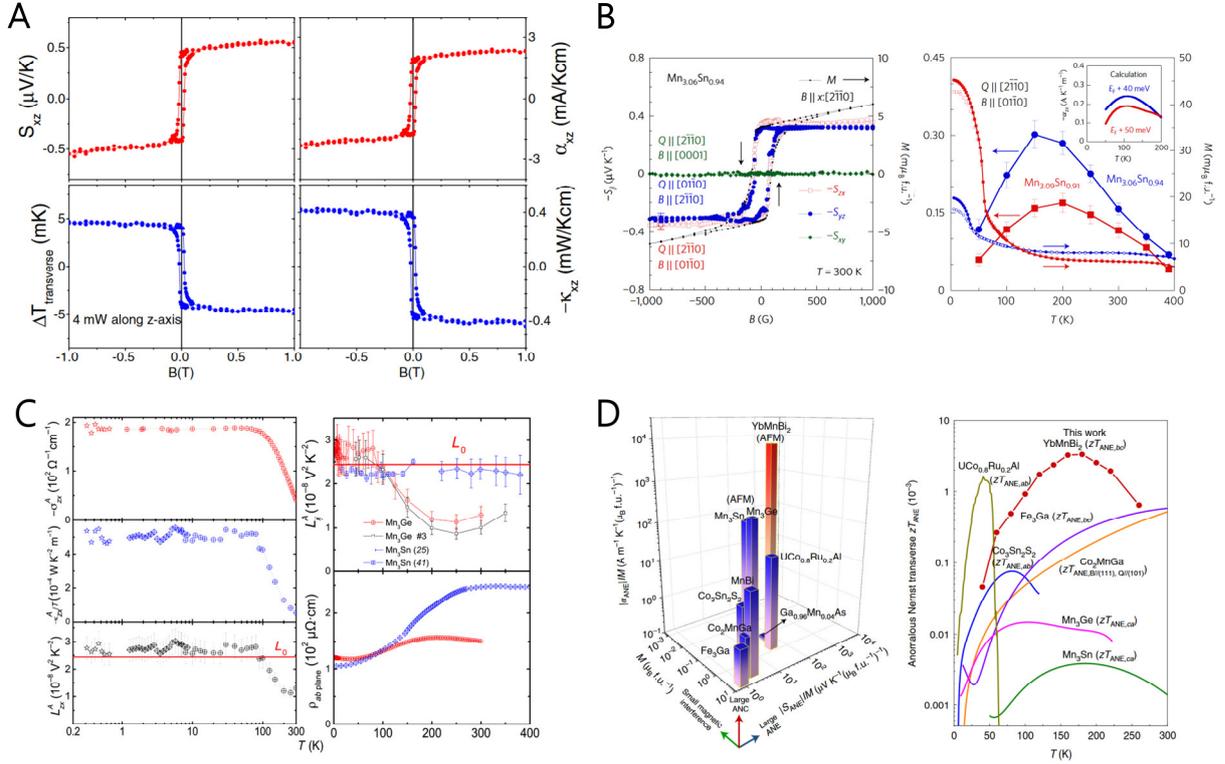

**Figure 10.** A) Nernst signal, transverse thermoelectric conductivity, transverse thermal gradient, and the extracted Righi-Leduc coefficient of Mn$_3$Sn as a function of the magnetic field. Reproduced under the terms of the CC-BY 4.0 license.[225] Copyright 2017, The Authors, Published by American Physical Society. B) Anisotropic field dependence of the Nernst signal and temperature dependence of transverse thermoelectric conductivity. Reproduced with permission.[226] Copyright 2017, Springer Nature. C) Left panel: temperature dependence of the AHC, ATHC divided by temperature, and anomalous Lorenz ratio in Mn$_3$Ge crystal. Right panel: Comparative temperature-dependent variations of anomalous Lorentz ratio and in-plane resistivity for Mn$_3$Ge and Mn$_3$Sn. Reproduced with permission.[229] Copyright 2020, American Association for the Advancement of Science. D) Comparison of the $|\alpha_{ANE}|/M$, $|S_{ANE}|/M$, M, and the temperature-dependent zT$_{ANE}$ values for YbMnBi$_2$ with those of other compounds, where $\alpha_{ANE}$ represents ANE conductivity, $S_{ANE}$ ANE thermopower, M small magnetization, zT$_{ANE}$ the high anomalous Nernst thermoelectric figure of merit. Reproduced with permission.[232] Copyright 2022, Springer Nature.

### 3.2.2 MOEs

Since the discovery of MOKE and MOFE over a century ago, two widely accepted principles have been established based on band theory: (a) MOEs scale with the spontaneous magnetization of the material. Consequently, AFMs, which exhibit zero net magnetization, were conventionally assumed to lack measurable MOE signals. (b) The physical origin of MOE lies in the coexistence of SOC and band exchange splitting, with the latter resulting from finite net magnetization. However, recent advances have challenged these long-standing interpretations, offering new insights into MOE. In 2015, Feng et al. first demonstrated significant MOKE in ncl-AFMs $Mn_3X$ ($X$=Ir, Pt, Rh) through first-principles calculations and symmetrical analysis.[237] As shown in **Figures 11A-B**, the nonzero MOEs are presented in both T1 and T2 magnetic configurations despite their vanishing net magnetization. The overall characteristics and peak values of MOEs follow the trend $Mn_3Pt > Mn_3Ir > Mn_3Rh$, suggesting that the strength of SOC in the $X$ atom plays a crucial role in enhancing the MOKE spectra. Notably, $Mn_3Pt$ exhibits a maximum Kerr rotation angle of approximately 0.6°, which is comparable to those observed in ferromagnetic transition metals.[238,239] These findings indicate that the band exchange splitting due to the broken $\mathcal{TS}$ symmetry, in combination with SOC, is responsible for the emergence of the MOEs in $Mn_3X$. Subsequently, Zhou et al. further investigated the effects of chiral- and spin-orders on MOEs in ncl-AFMs $Mn_3XN$ ($X$ = Ga, Zn, Ag, Ni), revealing the vector-spin chirality dependent MOEs.[173] Furthermore, Wimmer et al. systematically examined the MOKE and X-ray magnetic dichroism in the $Mn_3X$ ($X$ = Ga, Ge, Sn, Rh, Ir, Pt) family.[240]

The first experimental detection of MOKE in ncl-AFMs was achieved in $Mn_3Sn$, which exhibits a large zero-field Kerr rotation angle of 20 mdeg at room temperature (**Figure 11C**).[241] Notably, the Kerr effect remains largely unaffected by increasing external magnetic fields and magnetization, with the ratio 25.6 deg T$^{-1}$ of Kerr angle to magnetization reaching up to one to two orders of magnitude higher compared to FMs and conventional AFMs. This indicates that the MOKE amplitude is completely decoupled from the magnetization. MOKE microscopy was successfully employed to image magnetic domains in the noncollinear antiferromagnetic state of $Mn_3Sn$, demonstrating its effectiveness as a non-contact, and non-destructive technique for imaging magnetic domains and their associated dynamics (**Figure 11D**). Subsequently, Balk et al. and Wu et al. independently measured the magneto-optical Kerr signal in $Mn_3Sn$ and $Mn_3Ge$, respectively.[242,243]

In 2020, Feng et al. further proposed a novel MOE family member, termed topological MOE, which originates from finite scalar spin chirality and can emerge in noncoplanar AFMs without any reference to SOC and band exchange splitting, fundamentally differing from the conventional MOE mechanism in FMs, AMs, and coplanar ncl-AFMs.[244] As shown in **Figure 11E**, the topological MOEs appear in the distorted noncoplanar antiferromagnetic $\gamma$-$Fe_{0.5}Mn_{0.5}$ alloy with the 3Q state, while vanishing in the 1Q and 2Q states due to the zero scalar spin chirality. More intriguingly, in insulating topological ncl-AFMs, topological MOEs can be quantized, resulting in the quantized Kerr and Faraday rotation angles in the low-frequency limit, as demonstrated in the layered rhodium oxide $K_{0.5}RhO_2$ and monolayer $RhO_2$ (**Figure 11F**).[244] Beyond linear MOEs, scalar spin chirality in ncl-AFMs also plays a role in higher-order magneto-optical effects, such as nonlinear Kerr effect, Voigt phenomena, and magnetic dichroism. For example, Yang et al. recently predicted the second-order topological magneto-optical effects, including topological Voigt and Schäfer-Hubert effects.[245] Very recently, the topological MOKE was experimentally observed in Gd$_2$PbSi$_3$ and CrVI$_6$ (**Figures 11G-H**), both of which exhibit a skyrmion lattice structure.[246,247] These discoveries provide an innovative design strategy for developing magneto-optical devices, representing a significant departure from conventional approaches that rely on elements with strong SOC.

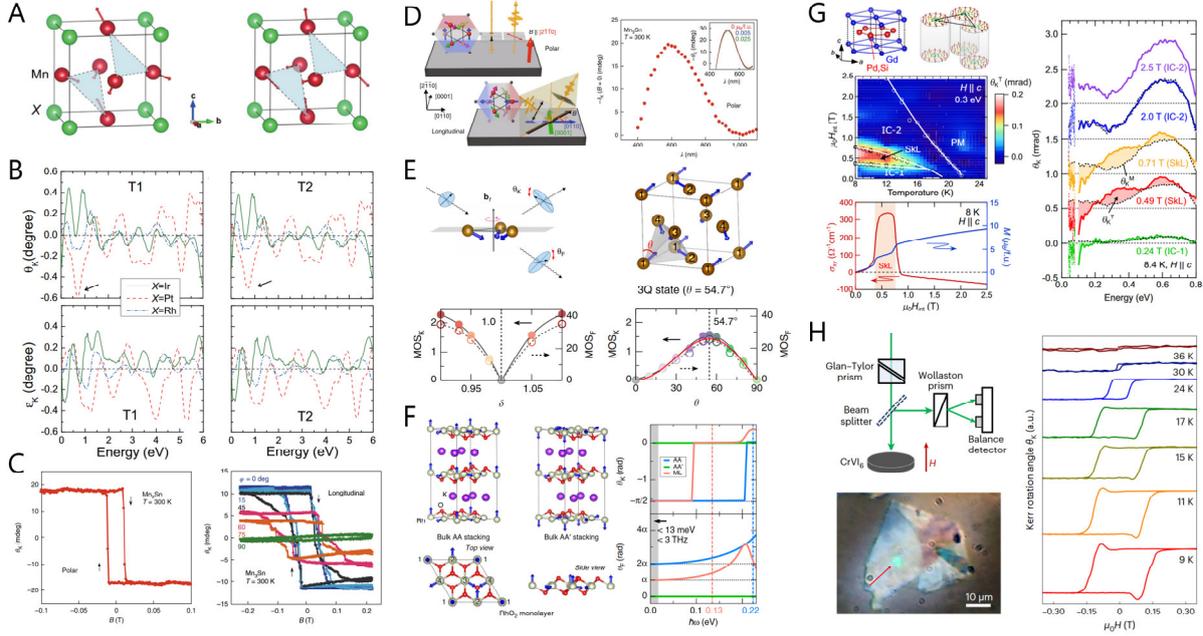

**Figure 11.** A) Crystal structure of Mn$_3$X (X = Rh, Ir, Pt) with T1 (left panel) and T2 (right panel) spin configurations. B) The Kerr rotations angles (upper panels) and Kerr ellipticities angles (lower panels) for the T1 and T2 spin configurations of Mn$_3$X alloys. Reproduced under the terms of the CC-BY 4.0 license.[237] Copyright 2015, The Authors, Published by American Physical Society. C) Polar magneto-optical Kerr rotation angle $\theta_K$ of the $(2\bar{1}\bar{1}0)$ plane under $B\|[2\bar{1}\bar{1}0]$ (left panel) and longitudinal Kerr rotation angle $\theta_K$ of the $(2\bar{1}\bar{1}0)$ plane measured along $\varphi = 0–90°$ (right panel). D) Left panels: Diagram of sample configurations for polar and longitudinal MOKE measurements. Right panels: Polar MOKE spectroscopy of Mn$_3$Sn crystal under zero magnetic field. Reproduced with permission.[241] Copyright 2018, Springer Nature. E) Top panels: schematic illustration of topological light-matter interactions in a minimal chiral magnet composed of three neighboring noncoplanar spins and 3Q state spin textures of $\gamma$-Fe$_x$Mn$_{1-x}$. Bottom panels: The strain (left) and $\theta$ (right) dependence of the magneto-optical strength of Kerr and Faraday effects in $\gamma$-Fe$_x$Mn$_{1-x}$ (3Q state). F) Crystal structures, magnetic structures, and the Kerr and Faraday rotation angles of K$_{0.5}$RhO$_2$ with AA and AA' stacking and monolayer RhO$_2$. Reproduced with permission.[244] Copyright 2020, Springer Nature. G) The crystal structure, triangular skyrmion lattice, magnetic phase diagram, magnetic-field dependence of the Hall conductivity, and magneto-optical Kerr rotation angle of Gd$_2$PdSi$_3$. Reproduced with permission.[246] Copyright 2023, Springer Nature. H) Left panels: Optical path for polar MOKE measurements and optical image of an exfoliated CrVI$_6$ thin film. Right panel: temperature-dependent hysteresis loops of the Kerr rotation angle, with the bumps signifying

the presence of the topological Kerr effect. Reproduced with permission.[247] Copyright 2024, Springer Nature.

## 4. Layer-polarized AFMs

When magnets transition from 3D to 2D systems, their magnetic properties undergo significant changes due to reduced symmetry, altered interactions, and enhanced fluctuations. In an ideal 2D isotropic Heisenberg system, long-range magnetic order at finite temperature is forbidden due to strong thermal fluctuations. However, real materials often have anisotropies (e.g., Ising-type or XY-type interaction), which can stabilize magnetic ordering even in the monolayer limit, as demonstrated in materials such as $CrX_3$ ($X$ = Cl, Br, I),[248–250] $Cr_2Ge_2Te_6$,[251] $Fe_3GeTe_2$,[252] and so on. Unlike conventional bulk magnets, the interlayer exchange interactions in these Van der Waals magnets can be tuned through sliding, rotation (twisting), and stacking modifications, allowing for external control over magnetic properties. This tunability opens pathways for controllable magnetism symmetry. Layer-polarized AFMs (LP-AFMs), represent a class of layered antiferromagnetic materials that lack $\mathcal{TS}$ symmetry, exhibiting layer-dependent spin polarization and magnetic transport properties. The layer physics, coupled with additional degrees of freedom in the 2D system, including valley-indexes, multiferroic order, non-dissipative topological transport, and axion electrodynamics, enable multidisciplinary applications in the nanoscience spintronics. Here, we take several well-studied LP-AFMs as prototypical examples to illustrate their mechanisms and applications.

### 4.1 Global spin-polarization

The exploration of 2D magnets lags behind 3D counterparts due to the practical challenges in synthesis, stability, and characterization. To date, most experimentally confirmed monolayer magnetic materials are FMs rather than AFMs. In contrast, even layer magnets, such as bilayer materials, often harbor antiferromagnetic order, typically in the A-type configuration, characterized by ferromagnetic ordering within layers and antiferromagnetic coupling between layers, as seen in bilayer $CrI_3$, $Cr_2Ge_2Te_6$, $Fe_3GeTe_2$, and $MnBi_2Te_4$. When a potential difference is introduced between AFM layers, the $\mathcal{TP}$ symmetry is broken, enabling the desired spin-splitting phenomena, which can be induced by layer stacking or electric field effects. As shown in **Figures 12A-B**, a honeycomb lattice with collinear Néel order, exemplified by bilayer $MnPSe_3$, exhibits spin degeneracy of electronic bands at the entire Brillouin zone, protected by $\mathcal{PT}$-symmetry. However, when a vertical electric field or an epitaxial substrate was applied, the $\mathcal{PT}$-symmetry was broken, resulting in spin-splitting of the bands despite the absence of net magnetization.[25] Notably, this spin degeneracy is lifted across the entire Brillouin zone, in contrast to the localized spin polarization observed in AMs.[253] In $\mathcal{PT}$-symmetric layered AFMs, although the net spin polarization remains zero, each individual layer exhibits hidden spin polarization, which is locked to the layer index and oppositely oriented in adjacent layers. The application of an electric field induces the spatial separation of layer-polarized currents, breaking the balance between layers and leading to novel global spin polarization effects. This layer spin polarization has been observed in bilayer $MnBi_2Te_4$,[26] which will be thoroughly reviewed in the following section. Very recently, Yao et al. constructed a dual-gate transistor using bilayer van der Waals semiconductor CrPS$_4$ with A-type antiferromagnetic order and demonstrated a switchable control of conduction band spin polarization via a perpendicular electric displacement field.[254] As shown in **Figures 12C-D**, at zero electric field, the energy bands of the top and bottom layers are degenerate, and time-reversal paired magnetic states A and B have equal energy, resulting in no hysteresis. When an electric field is applied, it breaks $\mathcal{PT}$-symmetry, lowering the energy of one layer's conduction band. The Zeeman energy difference between the two magnetic states leads to an energy splitting, where state A is favored under a positive magnetic field, and state B is favored under a negative magnetic field. As the magnetic field sweep direction reverses, the system switches between states A and B, giving rise to magnetoconductance hysteresis.

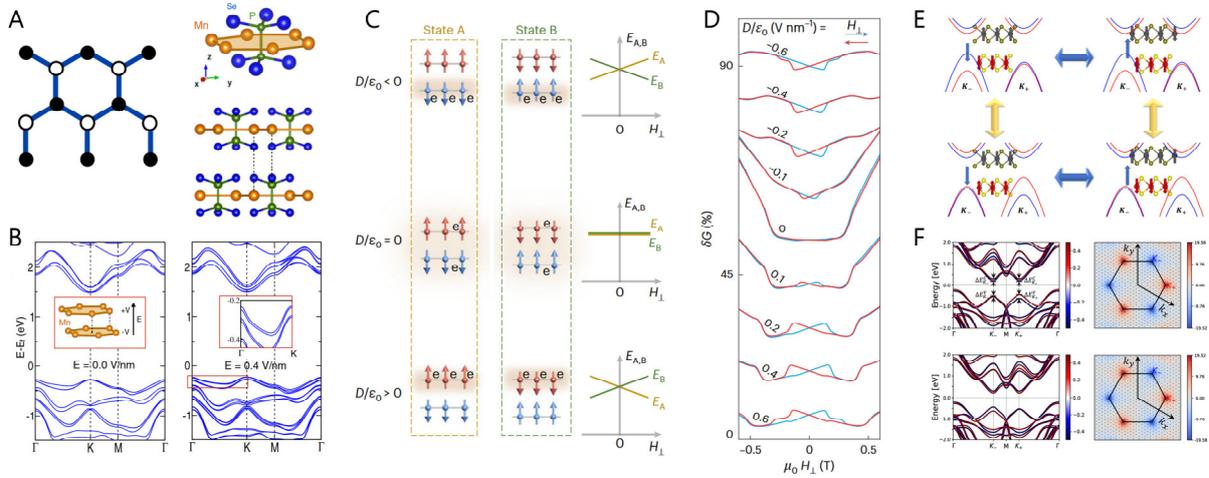

**Figure 12.** A) Schematic of a honeycomb lattice with collinear Néel order and the crystal structure of monolayers and bilayer MnPSe$_3$. B) The band structure of the bilayer MnPSe$_3$ without electric field (left) and with 0.4 V/nm electric field (right). Reproduced under the terms of the CC-BY 4.0 license.[25] Copyright 2016, The Authors, Published by American Physical Society. C) Degeneracy breaking of time-reversal paired states A and B of bilayer CrPS$_4$ due to spin-polarized electron accumulation under finite electric field. D) Magnetoconductance hysteresis as experimental evidence of energy differentiation between states A and B at different $D/\varepsilon_0$. Reproduced with permission.[254] Copyright 2025, Springer Nature. E) Schematic illustration of multistate control of 2D multiferroic bilayer VS$_2$. F) Band structure with SOC and Berry curvature distribution for different ferroelectric antiferromagnetic configurations (top panels: P↓M↑↓ and P↑M↓↑, bottom panels: P↓M↓↑ and P↑M↑↓. Reproduced under the terms of the CC-BY 4.0 license.[259] Copyright 2016, The Authors, Published by American Physical Society.

Building on these discoveries, researchers have extended the exploration of spin-layer physics to other layered materials, particularly those with A-type antiferromagnetic order, including bilayer MnBi$_2$Te$_4$,[26,27,255,256] bilayer CrI$_3$,[248,257,258] bilayer VS$_2$[259] and others.[28,260–272] The layer degree of freedom in 2D layered AFMs enables the breaking of $\mathcal{P}$ symmetry through stacking configuration modulation, thereby inducing both spin-layer polarization and sliding ferroelectricity.[259] As shown in **Figures 12E-F**, the 3R-type stacking of bilayer VS$_2$ breaks $\mathcal{P}$ symmetry, generating a spontaneous out-of-plane ferroelectric polarization. The coupling between ferroelectricity and A-type antiferromagnetism in bilayer VS$_2$ gives rise to spin-polarized band structures with asymmetric valley-dependent splittings and non-uniform Berry curvature distributions in the presence of SOC. Reversal of ferroelectric polarization (P↑↔P↓) synchronously flips the spin texture and Berry curvature sign, enabling electrically tunable valley-contrasting magnetic transport properties. Notably, in bilayer configurations where the two spin sublattices cannot be interconnected by any symmetry operations, the system exhibits a zero net magnetic moment while demonstrating spin band splitting across the entire Brillouin zone, even in the absence of SOC.[265] In conclusion, multilayer-stacked compensated AFMs manifest unique layer-polarized properties and ferroelectricity, whose coupling establishes them as promising candidates for high-density spintronic device platforms, offering exceptional potential for next-generation electronic integration.

**4.2. Layer hall effect**

The introduction of layer degrees of freedom has significantly advanced research on the study of the AHE. Through the coupling of this degree of freedom with antiferromagnetism, full electric-field control of the anomalous valley Hall effect has been achieved in LP-AFMs.[260] Specifically, in the initial state, antiferrovalley coupling preserves $\mathcal{PT}$ symmetry, maintaining valley degeneracy between the energy bands of the upper and lower layers and suppressing the AHE. However, applying a vertical electric field creates an interlayer potential difference that breaks $\mathcal{PT}$ symmetry, leading to band decoupling and valley energy splitting. A positive electric field shifts the lower layer subbands upward relative to the upper monolayer, whereas a negative electric field causes the opposite effect. Therefore, opposite electric fields generate Berry curvatures and Hall signals with opposite signs. In 2021, Gao et al. introduced the concept of the layer Hall effect to describe the AHE in even-layer MnBi$_2$Te$_4$.[26] In terms of symmetry, the essence of the layer Hall effect lies in the breaking of $\mathcal{PT}$ symmetry induced by the electric field. In A-type AFMs with $\mathcal{PT}$ symmetry, the spin structure is coupled with an even number of layers, resulting in the Dirac fermions on the top and bottom surfaces being magnetized in opposite directions. This leads to Berry curvatures of equal magnitude but opposite signs on the two layers. As a result, the anomalous Hall currents contributed by the top and bottom surfaces are equal in magnitude yet flow in opposite directions, canceling each other out overall (**Figure 13A**). However, when a positive out-of-plane electric field that breaks $\mathcal{PT}$ symmetry is applied, the Berry curvature from the bottom layer becomes dominant, producing a finite layer-polarized AHE. Conversely, when a negative out-of-plane electric field is applied, the top layer's Berry curvature dominates, causing the sign of the AHE to reverse (**Figure 13B**). The layer Hall effect was successfully captured in a well-designed dual-gated six-layer MnBi$_2$Te$_4$ devices (**Figure 13A**). Experimental results demonstrate that an out-of-plane electric field induces a layer-polarized AHC of approximately 0.5 e²/h in six-layer MnBi$_2$Te$_4$, with sign reversal upon electric field inversion (**Figure 13C**). Furthermore, Chen et al. demonstrated that the electric field mainly functions to separate the hidden Berry curvatures of different MnBi$_2$Te$_4$ layers, which are compensated by $\mathcal{PT}$ symmetry, through a comparison of the Hall conductivities between A-type antiferromagnetic topological insulators and trivial insulators under identical electric fields.[256] Besides electric field, $\mathcal{PT}$ symmetry breaking can also be achieved through modification of stacking configurations or by forming heterostructures with non-magnetic materials, providing alternative methods for realizing the layer Hall effect.[261,262,265–267,269–271,273] Intriguingly, interlayer ferroelectric polarization arising from sliding ferroelectricity offers a switchable polarization control over the sign of the layer Hall effect.[27,255,263,264,268,274,275] As shown in **Figure 13D**, the polarized stacking configuration breaks $\mathcal{P}$ symmetry, thereby leading to the breaking of $\mathcal{PT}$ symmetry. This results in a non-zero Berry curvature and layer Hall effect in bilayer MnBi$_2$Te$_4$. The two polarized phases are related by $\mathcal{T}M_z$ symmetry, which leads to opposite signs of Berry curvature and AHC in the two states ($\mathcal{T}M_Z\Omega_{xy}(k) = -\Omega_{xy}(-k)$).[27] In addition to the layer Hall effect, the layer Nernst effect and the layer thermal Hall effect, induced by out-of-plane electric fields breaking $\mathcal{PT}$ symmetry, have also been predicted in bilayer A-type antiferromagnetic systems such as Hf$_2$S, MnBi$_2$Te$_4$, CrI$_3$, and VS$_2$ (**Figures 13F-G**).[276,277] In summary, $\mathcal{PT}$ symmetry breaking is the fundamental mechanism underlying the layer transport effects in LP-AFMs. These low-dimensional, easily tunable 2D LP-AFMs offer an ideal platform for the development of highly integrated spintronic devices. Their intrinsic sensitivity to external stimuli, such as electric fields, stacking configurations, and interlayer coupling, paves the way for next-generation spintronic applications with enhanced functionality, energy efficiency, and scalability.

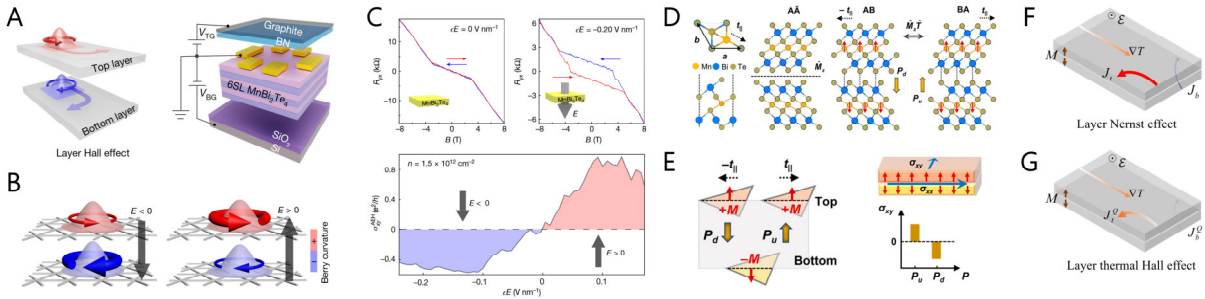

**Figure 13.** A) Illustration of the layer Hall effect and schematic drawing of dual-gated devices. B) Illustration of the layer-locked Berry curvature under opposite electric fields. C) Top panels: the Hall resistance as $R_{yx}$ dependence of magnetic field. Bottom panels: The AHE conductivity $\sigma_{xy}^{AHE}$ as a function of an electric field at zero magnetic field. Reproduced with permission. Copyright 2021,[26] Springer Nature. D) Atomic structure of monolayer MnBi$_2$Te$_4$, the unstable A$\overline{\text{A}}$ stacking that spontaneously relaxes to the lower-energy ferroelectric phase with AB and BA stacking. E) Schematic illustration of the tunable AHE in bilayer MnBi$_2$Te$_4$ with polar stacking. Reproduced with permission.[27] Copyright 2021, American Chemical Society. F-G) Schematic illustration of the layer Nernst and layer thermal Hall effects. Reproduced under the terms of the CC-BY 4.0 license.[277] Copyright 2025, The Authors, Published by American Physical Society.

4.3 MOEs

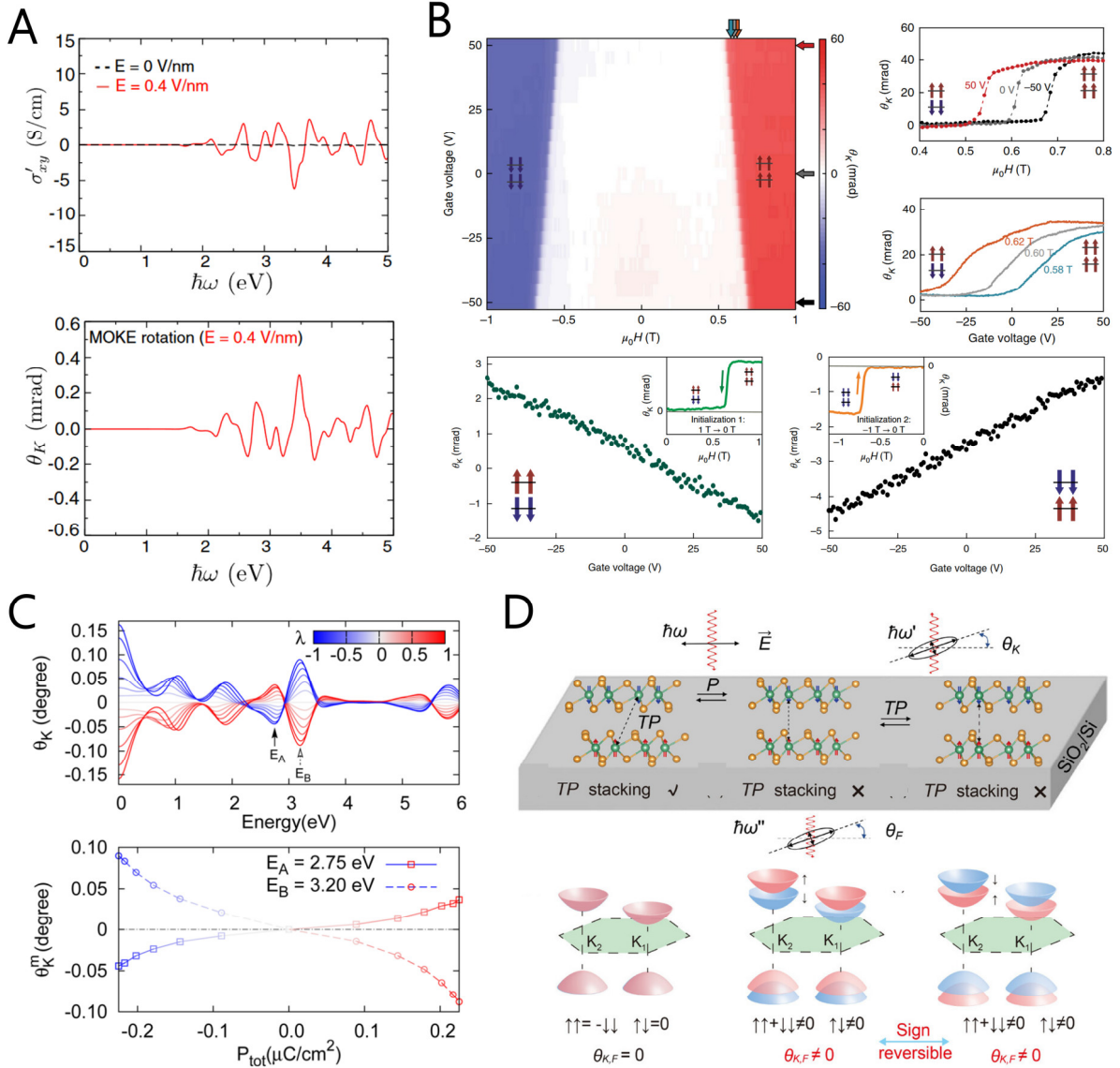

**Figure 14.** A) The optical Hall conductivity and Kerr angle of bilayer MnPSe$_3$ at finite electric field Reproduced under the terms of the CC-BY 4.0 license.[25] Copyright 2016, The Authors, Published by American Physical Society. B) Top panels: Intensity of the polar MOKE signal of bilayer CrI$_3$ under different magnetic fields and different gate voltages. Bottom panels: gate-voltage-induced MOKE of bilayer CrI$_3$ at zero magnetic field. Reproduced with permission.[258] Copyright 2018, Springer Nature. C) Top panels: Kerr rotation dependence of the incident photon energy, λ represents the normalized amplitude of atomic distortion. Bottom panels: Some maxima and minimum Kerr rotation are shown as a function of ferroelectric polarization. Reproduced with permission.[278] Copyright 2017, American Chemical Society. D) Schematic illustration of MOE and spin-valley-polarized property with different stacking configurations of bilayer Nb$_3$I$_8$. Reproduced with permission.[28] Copyright 2024, American Chemical Society.

Similar to the layer hall effect, breaking $\mathcal{PT}$ symmetry can also induce MOEs in LP-AFMs. In 2016, Sivadas et al. first elucidated this mechanism through systematic theoretical analysis and model construction, revealing that the emergence of MOEs fundamentally stems from the breaking of magnetic symmetries ($\mathcal{PT}, \mathcal{T}C_{2z}, \mathcal{T}M_z$).[25] As shown in **Figure 14A**, in bilayer MnPSe$_3$ with collinear compensated antiferromagnetic order, the optical conductivity tensor $\sigma_{xy}$ remains zero in the absence of an electric field constrained by $\mathcal{PT}$ symmetry. Upon applying a small electric field (0.4 V/nm) to break $\mathcal{PT}$ symmetry, $\sigma_{xy}$ becomes nonzero, accompanied by a detectable polar Kerr angle of 0.3 mrad. Notably, monolayer MnPSe$_3$, which shares the collinear compensated antiferromagnetic order and $\mathcal{PT}$ symmetry, exhibits a vanishingly small Kerr angle under identical electric field conditions. This distinction arises because $\mathcal{P}$-breaking in the monolayer is achieved through a potential difference between the top and bottom PSe$_3$ layers, which affects Mn atoms via weak interactions between Mn *d*-orbitals and Se *p*-orbitals. In contrast, Mn atoms in different layers of the bilayer system directly experience the applied electric field, resulting in significant layer-polarization and stronger symmetry-breaking effects. Utilizing magneto-optical effects, bilayer CrI$_3$ was experimentally confirmed to host intralayer ferromagnetic and interlayer antiferromagnetic ground states through magnetic circular dichroism microscopy.[257] This finding also underscores the high potential of 2D AFMs as a versatile platform for achieving voltage-tunable magneto-optical functionalities. As shown in **Figure 14B**, the full electrical control of antiferromagnetic states has been realized in bilayer CrI$_3$ probed by MOKE microscopy.[258] At zero magnetic field, the MOKE signals of the time-reversal pairs of layered antiferromagnetic states exhibit a linear dependence on the back-gate voltage with opposite slopes, confirming the spin-layer locking mechanism in LP-AFMs.

In addition to applying an electric field, the antiferromagnetic system with ferroelectricity, like a hybrid organic−inorganic perovskite Cr-metal-organic framework, can also exhibit MOEs, due to the breaking of $\mathcal{PT}$ symmetry induced by its ferroelectricity and collinear antiferromagnetic order.[278] Moreover, along the ferroelectric phase transition path, the magneto-optical Kerr angle increases with the enhancement of polarization strength and reverses its sign upon ferroelectric polarization switching (**Figure 14C**). The MOEs in AFMs can also be active by heterojunction engineering, as demonstrated in CrBr$_3$/CrI$_3$[279] and CrI$_3$/In$_2$Se$_3$/CrI$_3$.[280] Additionally, layer twisting also provides an effective strategy for manipulating magnetic symmetries and MOKE. The Kerr angle is controllable by the twist angle, and its sign is reversed when switching between left- and right-twisted bilayers.[281]

Recently, Zhou et al. provided a novel method for controlling MOEs through different layer-stacking configurations.[28] As shown in **Figure 14D**, in the bilayer Nb$_3$I$_8$ system, by adjusting the layer stacking order, the spin splitting and valley splitting in bilayer Nb$_3$I$_8$ can be dynamically tuned, enabling the on/off switching of the magneto-optical signal. In conventional $\mathcal{PT}$-stacked structures, valley polarization exists; however, due to the spin degeneracy of the energy bands, optical transitions between the two $\mathcal{PT}$-symmetric sets of energy bands cancel each other out, leading to the disappearance of the MOEs. In contrast, layer stacking that breaks $\mathcal{PT}$ symmetry induces simultaneous spin and valley splitting. Unlike the spin-momentum locking feature in AMs, this system exhibits spin splitting across the entire Brillouin zone. This unique spin-polarization feature, combined with strong spin-orbit coupling, enables spin-flip interband transitions, giving rise to nonzero MOEs. A distinct characteristic of the layer-controlled magneto-optical effect is that its sign is jointly governed by the ferroelectric polarization direction and the Néel vector orientation. In topological antiferromagnet $MnBi_2Te_4$, this reversible magneto-optical effect can even become quantized in the low-frequency limit, leading to a quantum magneto-optical effect. In summary, in 2D LP-AFMs, MOEs can be effectively tuned by modifying the layer-stacking sequence, applying external electric fields, or reversing ferroelectric polarization.[272,279–284] These strategies allow for the precise control of the material's magnetic and optical properties. Remarkably, significant magneto-optical responses can be observed even in the absence of an external magnetic field, highlighting the potential of 2D LP-AFMs for future applications in magnetic and optoelectronic devices where field-independent control of magneto-optical properties is desirable.

## 5. Summary and Outlooks

This review systematically examines the spin polarization and magnetic transport properties in AFMs, with a particular focus on the symmetry-breaking mechanisms (e.g., rotational sublattice coupling, chiral spin textures, and layer-polarized potentials) responsible for these phenomena in AMs, ncl-AFMs, and LP-AFMs. Unlike traditional AFMs, these antiferromagnetic systems exhibit ferromagnet-like transport properties without the need for external magnetic fields, while avoiding magnetic interference in highly integrated devices, offering significant practical advantages. Furthermore, they provide promising opportunities for efficient manipulation of transport and optical responses through symmetry-engineered band topology, spin chirality, and interlayer reconstruction, which not only deepen our understanding of the fundamental physics of spin-polarized antiferromagnetic spintronics but also open new venues for the design of high-density, energy-efficient spin-based electronic devices.

Despite the rapid progress in this field in recent years, several key scientific challenges remain to be addressed to facilitate the practical application of these materials in functional devices: i) The number of materials exhibiting room-temperature antiferromagnetic order with robust spin-polarized transport and magneto-optical responses remains limited. Addressing this requires high-throughput computational screening guided by magnetic symmetric principles, coupled with advanced synthesis techniques such as strain-stabilized epitaxy and defect-passivated 2D growth to enhance dimensional stability and preserve spin polarization in low-dimensional systems. ii) While many spin-polarized effects depend on low temperatures, multifield approaches-combining electric fields, strain gradients, and interfacial charge transfer- could enable tunable Néel vector control at ambient conditions. Interlayer stacking engineering and hybrid heterostructure (e.g., AFM/ferroelectric interfaces) further offer pathways to achieve dynamic symmetry breaking for industrial-scale applications. iii) Topological states (e.g., Weyl nodes, nodal lines) intrinsic to spin-polarized AFMs provide a platform to amplify Berry curvature effects and stabilize dissipationless spin currents. Future efforts should prioritize designing topology-protected spin valves and non-local circuits to minimize energy consumption and enhance signal integrity in spin information transmission. iv) Leveraging

THz-scale dynamics and stray-field immunity, spin-polarized AFMs are ideal candidates for ultrahigh-density memory, non-von Neumann logic architectures, and quantum coherent interconnects. Key innovations include integrating AFMs with SOT modules and topological electronics to bridge ultrafast antiferromagnetic switching with Complementary Metal-Oxide-Semiconductor-compatible manufacturing. In conclusion, spin-polarized antiferromagnetic materials have overcome the limitations of traditional magnetic materials and opened new avenues in spintronics. Through synergistic advancements in symmetry-driven material discovery, multifield spin control, and topology-enhanced device engineering, these systems are poised to drive breakthroughs in energy-efficient computing, secure quantum communication, and post-Moore's law technologies.

Acknowledgements

This work was supported by the National Key R&D Program of China (Grant No. 2022YFA1402600) and the Australian Research Council Discovery Early Career Researcher Award (Grant No. DE240100627).

Received: ((will be filled in by the editorial staff))
Revised: ((will be filled in by the editorial staff))
Published online: ((will be filled in by the editorial staff))